\documentclass[a4paper,11pt]{article}
\usepackage{jcappub} 

\usepackage[T1]{fontenc} 

\usepackage{graphicx, epsfig, amssymb} 
\usepackage{amsmath, amsfonts}
\usepackage{bm} 

\def\be{\begin{equation}}
\def\ee{\end{equation}}
\def\beq{\begin{eqnarray}}
\def\eeq{\end{eqnarray}}

\title{\boldmath Constraining the mass of dark photons and axion-like particles through black-hole superradiance}

\author{
Vitor Cardoso$^{1,2}$,
\'Oscar J. C. Dias$^{3}$,
Gavin S. Hartnett$^{3}$,
Matthew Middleton$^{4}$,
Paolo Pani$^{5}$,
Jorge E. Santos$^{6}$
}
\affiliation{${^1}$ CENTRA, Departamento de F\'{\i}sica, Instituto Superior T\'ecnico -- IST, Universidade de Lisboa -- UL,
Avenida Rovisco Pais 1, 1049 Lisboa, Portugal}
\affiliation{${^2}$ Perimeter Institute for Theoretical Physics, 31 Caroline Street North
Waterloo, Ontario N2L 2Y5, Canada}
\affiliation{${^3}$ STAG research centre and Mathematical Sciences, University of Southampton, UK}
\affiliation{${^4}$ Department of Physics and Astronomy, University of Southampton, Highfield, Southampton SO17 1BJ, UK}
\affiliation{${^5}$ Dipartimento di Fisica, ``Sapienza'' Universit\`a di Roma \& Sezione INFN Roma1, Piazzale Aldo Moro 5, 00185, Roma, Italy}
\affiliation{${^6}$ DAMTP, Centre for Mathematical Sciences, University of Cambridge, Wilberforce Road, Cambridge CB3 0WA, UK}

\abstract{
Ultralight bosons and axion-like particles appear naturally in different scenarios and could solve some long-standing puzzles. Their detection is challenging, and all direct methods
hinge on unknown couplings to the Standard Model of particle physics. 
However, the universal coupling to gravity provides model-independent signatures for these fields. We explore here the superradiant instability
of spinning black holes triggered in the presence of such fields. The instability taps angular momentum from and limits the maximum spin of astrophysical black holes.
We compute, for the first time, the spectrum of the most unstable modes of a massive vector (Proca) field for generic black-hole spin and Proca mass. 
The observed stability of the inner disk of stellar-mass black holes can be used to derive \emph{direct} constraints on the mass of dark photons in the mass range $ 10^{-13}\,{\rm eV}\lesssim m_V \lesssim 3\times 10^{-12}\,{\rm eV}$. By including also higher azimuthal modes, similar constraints apply to axion-like particles in the mass range $6\times10^{-13}\,{\rm eV}\lesssim m_{\rm ALP} \lesssim 10^{-11}\, {\rm eV}$.
Likewise, mass and spin distributions of supermassive BHs --~as measured through continuum fitting, K$\alpha$ iron line, or with the future space-based gravitational-wave detector LISA~-- imply indirect bounds in the mass range approximately $10^{-19}\,{\rm eV}\lesssim m_V, m_{\rm ALP} \lesssim  10^{-13}\, {\rm eV}$, for both axion-like particles and dark photons. Overall, superradiance allows to explore a region of approximately $8$ orders of magnitude in the mass of ultralight bosons.
}

\begin{document}
\maketitle
\flushbottom
\section{Introduction}

Two fundamental scales in nature --~set by the electron mass and by the radius of the Universe~--  are separated by roughly 40 orders of magnitude difference. 
New fundamental scales, such as those dictated by new interactions, could be expected to set in and populate this seemingly forlorn arena. 
In fact, there are arguments indicating that new, ultralight fields whose mass falls in this broad range, are natural outcomes which can simultaneously solve
important, long-standing puzzles.
Light bosonic fields with masses $\ll{\rm eV}$ are promising dark-matter candidates which offer a dramatically different phenomenology compared to that of weakly-interacting massive particles at the GeV-TeV scale. Furthermore, relativistic, ultralight boson fields can form macroscopic Bose-Einstein condensates which provide a natural alternative to the standard structure formation through dark-matter seeds and to the cold dark-matter paradigm (cf.~\cite{Suarez:2013iw,Li:2013nal,Hui:2016ltb} and references therein) .

The prototypical example of light bosons is the \emph{axion}, a light pseudoscalar which was introduced to solve the strong CP problem of QCD~\cite{Peccei:1977hh}. In this model, instanton effects give axions a tiny mass which is inversely proportional to the axion decay constant. By relying on this inverse proportionality, current bounds on QCD axions are derived from their overproduction in the early Universe, from negative searches in accelerators, and from astrophysical constraints~\cite{BertoneBook,Marsh:2015xka}.
%
%
\emph{Axion-like particles} (ALPs)~\cite{Jaeckel:2010ni,Essig:2013lka} have properties similar to those of the QCD axion but their mass is not related to the decay constant. Phenomenological constraints on these models are therefore much less stringent. These particles are ubiquitous in string-inspired scenarios such as the \emph{axiverse} and have been suggested as a generic signature of extra dimensions~\cite{Arvanitaki:2009fg,Acharya:2015zfk}. Due to moduli compactifications, in the axiverse scenario the spectrum of ALPs is populated uniformly down to the Hubble scale, $m_H\sim10^{-33}\,{\rm eV}$, so that ultralight bosonic states are allowed.
It has been recently recognized that ultralight boson fields with masses of the order of $10^{-21}\,{\rm eV}$ are a compelling candidate for cold dark matter~\cite{Hui:2016ltb}.
A similar, albeit much wider, phenomenology arises in models of \emph{ultralight vector fields~(ULVs)}, such as \emph{dark photons}, also a generic prediction of string theory~\cite{Goodsell:2009xc}. In this scenario, a ``hidden $U(1)$'' sector is weakly coupled to the visible Maxwell field through a kinetic mixing term.

\begin{figure}[th]
\begin{center}
\includegraphics[width=0.42\textwidth]{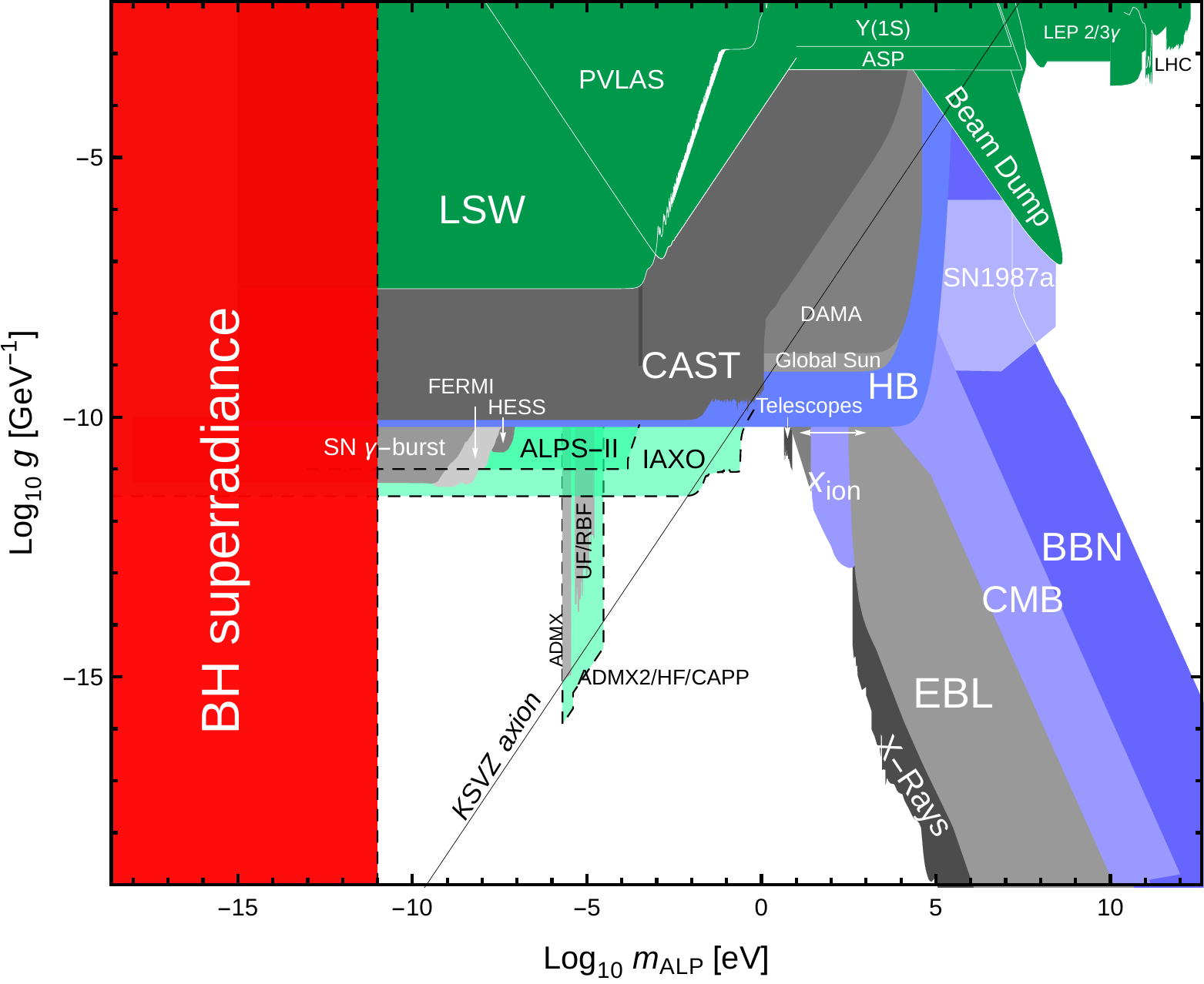}
\includegraphics[width=0.54\textwidth]{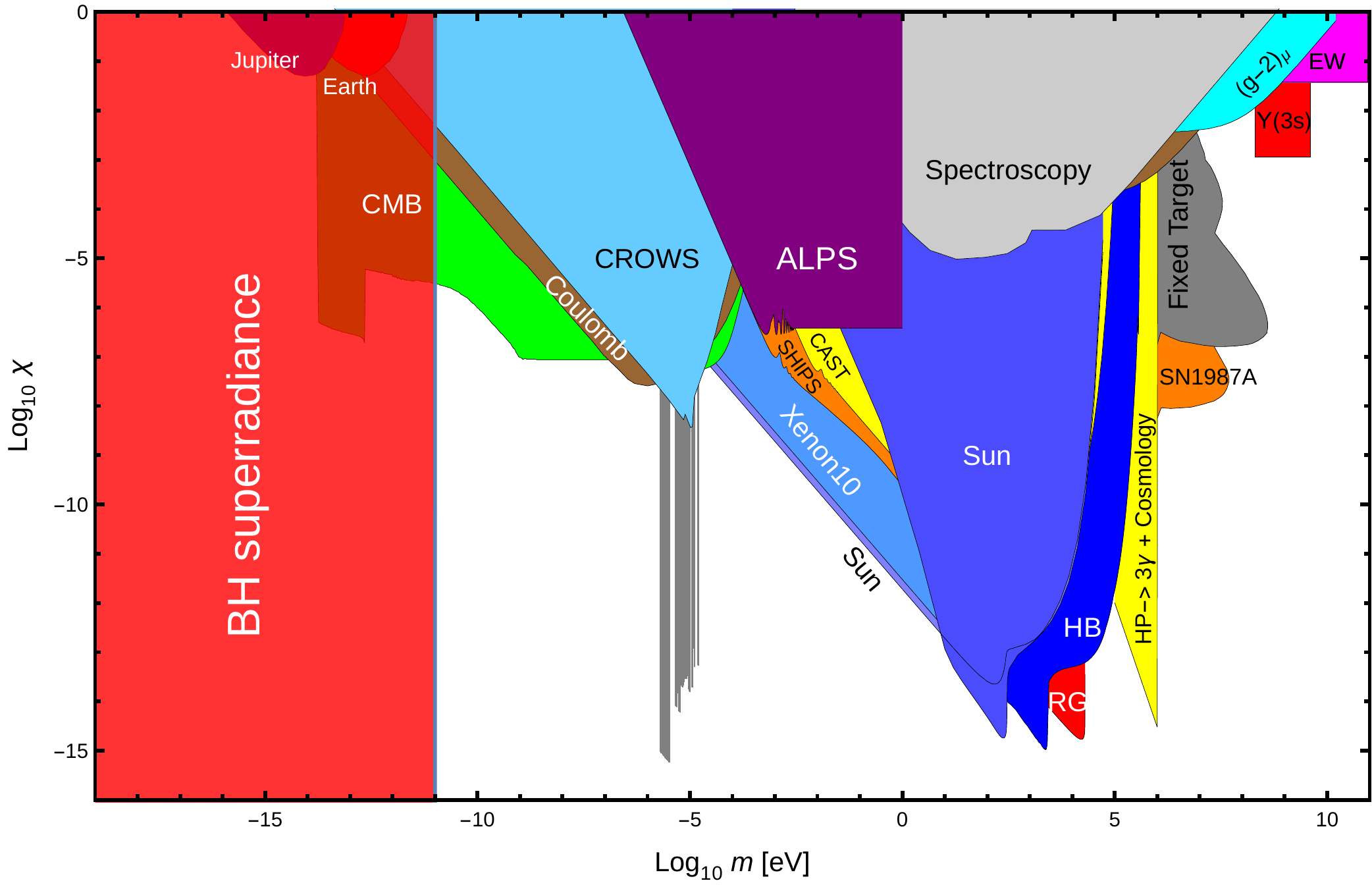}
\caption{Current experimental limits on ALPs (left) and ULVs (right) in their corresponding mass-coupling plane (adapted from~\cite{Sigl:2017wfx} and~\cite{Jaeckel:2013ija}, respectively; courtesy of J.\ Redondo). The red dashed areas denote the regions that can be probed through the superradiant instability of astrophysical BHs~\cite{Arvanitaki:2010sy,Pani:2012vp,Arvanitaki:2014wva,Baryakhtar:2017ngi,Brito:2017wnc,Brito:2017zvb} (cf.~\cite{Brito:2015oca} for an overview), as discussed in this paper. These constraints do not require a direct coupling between dark matter and ordinary particles, and are complementary to other bounds.
\label{fig:boundsbosons}}
\end{center}
\end{figure}
%

\noindent{\bf{\em Constraints on ultralight bosons.}}
%
Due to their tiny mass and weak coupling, direct searches of ultralight bosons in the lab are extremely challenging, especially for masses $\ll 10^{-10}\,{\rm eV}$.
The current bounds and respective experiments/observations from which they were derived are summarized in Fig.~\ref{fig:boundsbosons}. Laboratory searches inevitably require the interaction cross section of dark-matter particles with ``ordinary'' matter to be sufficiently large. For vanishingly small couplings to the Standard Model, gravity is the only interaction able to probe new fundamental fields.

There are at least two nontrivial effects of gravity on ALPs and ULVs. The first is the formation of self-gravitating structures such as boson stars or oscillatons~\cite{Liebling:2012fv}. These structures can become very compact, to the point of providing a compelling alternative to supermassive dark objects~\cite{Cardoso:2017cfl,Cardoso:2016oxy,Sennett:2017etc,Vincent:2015xta}.
The second, nonperturbative, effect is the triggering of \emph{superradiant instabilities} of spinning black holes (BHs)~\cite{Brito:2015oca}. Superradiant instabilities spin BHs down, and can affect the dynamics of astrophysical BHs in a dramatic fashion~\cite{Arvanitaki:2010sy,Brito:2015oca}, providing a portal for astrophysical tests of bosonic dark matter in the poorly explored range below $10^{-10}\,{\rm eV}$. 

\noindent{\bf{\em Superradiant instabilities of Kerr BHs and strong-gravity constraints.}}
The superradiant instability of ALPs and ULVs around spinning Kerr BHs is the focus of this work. 
The precise evolution or end-state of the instability for minimally coupled bosons is not fully understood~\cite{Brito:2015oca,Okawa:2014nda,Zilhao:2015tya},
but recent numerical simulations~\cite{East:2017ovw,East:2017mrj} support the
conclusions of previous perturbative studies~\cite{Brito:2014wla,Brito:2015oca}: 
the instability proceeds in a two-step process. During the first stage,
the geometry is well described by a (vacuum) Kerr BH. This geometry is unstable against nonaxisymmetric modes of massive bosonic fields, and any initial (low-frequency) small
fluctuation will grow exponentially~\cite{Detweiler:1980uk,Zouros:1979iw,Cardoso:2005vk,Dolan:2007mj,Brito:2015oca}. The boson accumulates outside the horizon,
forming an asymmetric (mostly dipolar in pattern) time-varying condensate, extending for roughly a de Broglie wavelength outside the horizon. Because of its purely gravitational nature, this effect is essentially independent of the dark-matter coupling to the Standard Model, provided the latter is sufficiently weak in order to avoid boson decay into other channels~\cite{Arvanitaki:2010sy}.

During the second stage, the condensate is massive enough that backreaction effects are no longer negligible.\footnote{For complex massive scalar fields in certain configurations, an interesting analytical model for the modified Kerr geometry was recently discussed in~\cite{Herdeiro:2017phl}.
Note, however, that the stationary hairy black holes~\cite{Herdeiro:2014goa} that exist in this case above the superradiant threshold for a given azimuthal quantum number $m$, were recently found to be unstable against nonaxisymmetric perturbations with azimuthal number larger than $m$~\cite{Ganchev:2017uuo}.} In particular,
the condensate emits monochromatic gravitational waves on large timescales. The condensate is then dissipated through the emission
of mostly quadrupolar gravitational waves, with frequency scale set by the boson mass.
The mechanism is most effective when the boson Compton wavelength is
comparable to the gravitational radius of the BH, $M\mu\simeq 0.4$ (henceforth, we use $G=c=1$ units and $\mu\hbar$ will denote the mass of the boson). Detailed calculations for the instability rate for scalar fields was done in~\cite{Cardoso:2005vk,Dolan:2007mj} and highly accurate numerical results will be given in Section~\ref{sec:spectrumscalar}.
Eventually, the BH spins down, the condensate shrinks and the emission of waves is suppressed~\cite{Brito:2014wla}.

Thus, strong-field gravity provides some novel mechanisms able to constraint ultralight bosons:
\begin{itemize}

\item {\bf Gravitational-wave emission by the bosonic condensate}. This is a monochromatic signal at a frequency $\sim \mu/\pi$ that can be detected with Earth- or space-based detectors, either as a resolvable
event or as part of a stochastic background. Recent calculations show that both would be seen by LIGO and LISA~\cite{Arvanitaki:2014wva,Arvanitaki:2016qwi,Baryakhtar:2017ngi,Brito:2017wnc,Brito:2017zvb}. In particular, the non-detection of these events in LIGO O1 could be used to exclude
certain mass ranges for the boson~\cite{Brito:2017wnc,Brito:2017zvb}.

\item {\bf Spin-down of astrophysical BHs and pulsars.} Gravitational-wave emission of periodic signal is a very clear sign of new physics. On the other hand, such an emission is driven by energy extraction from the BH, which spins down. Thus, another clear sign for new physics is statistical evidence for slowly rotating BHs in a part of the Regge plane (mass versus angular momentum plane)~\cite{Arvanitaki:2010sy,Brito:2015oca,Baryakhtar:2017ngi,Brito:2017wnc,Brito:2017zvb}. The X-ray spectrum of accreting compact sources, together with modeling of the emission from the accretion disc, provide a mean to measure the spin of stellar BHs in binaries and of active galactic nuclei~\cite{Middleton:2015osa}. These measurements are affected by the systematic uncertainty within the emission models. On the other hand, precise gravitational-wave measurements of the inspiral phase of two merging objects can provide accurate estimates for their spins~\cite{Arvanitaki:2014wva,Arvanitaki:2016qwi,Baryakhtar:2017ngi,Brito:2017wnc,Brito:2017zvb}. 
Likewise, superradiant instabilities might also occur in the presence of spinning stars made of material with nonvanishing resistivity~\cite{Cardoso:2017kgn}. In such case, the astonishing precision of pulsar timing provides accurate measurements of the spin and of the spin-down rate of pulsars, which can be used to set direct constraints on certain models of ULVs.
\end{itemize}

In summary,  electromagnetic and gravitational-wave observations provide a way to explore ultralight bosons in a mass range set by the size of astrophysical compact objects. Since the latter ranges from a few kilometers for stellar objects to ${\cal O}(10^9)\,{\rm km}$ for supermassive BHs, these observations can potentially cover (at least) 9 orders of magnitude. The potential constraints are mostly independent on the couplings of these fields to the Standard Model and are summarized in Fig.~\ref{fig:boundsbosons}, anticipating the discussion presented in the rest of this paper.

\noindent{\bf{\em Purpose of this work.}}
To be able to derive bounds on the boson mass, one needs to have accurate knowledge of how the superradiant instability proceeds in the two stages discussed above.
For scalars, the linearized instability regime is very well understood, both from frequency-domain calculations~\cite{Detweiler:1980uk,Zouros:1979iw,Cardoso:2005vk,Dolan:2007mj} and from time-domain evolutions~\cite{Dolan:2012yt}.
Nonlinear evolutions are still in their infancy~\cite{Okawa:2014nda}, but as we remarked it is thought that backreaction can be included using a perturbative approach~\cite{Yoshino:2013ofa,Brito:2014wla,Brito:2017wnc}.
Building on these analysis, bounds from gravitational radiation and from the Regge plane distribution were recently derived~\cite{Arvanitaki:2016qwi,Brito:2017wnc,Brito:2017zvb}.

The extension of the above results to massive vectors is nontrivial. In particular, the corresponding linearized evolution equations do not seem to separate~\cite{Rosa:2011my,Pani:2012vp,Pani:2012bp}, which makes the analysis somewhat more complicated.
A slow-rotation expansion was used in Refs.~\cite{Pani:2012vp,Pani:2012bp,Endlich:2016jgc}, and it was observed that the instability timescales could be substantially shorter~\cite{Pani:2012vp,Pani:2012bp} than for scalars (thus potentially improving constraints on ULVs). Time-domain, linearized investigations extended these predictions to highly spinning BHs~\cite{Witek:2012tr}.
Analytical results valid for small coupling parameter $M\mu$ but for arbitrary spin were recently derived in Ref.~\cite{Baryakhtar:2017ngi}. 
Very recently, full nonlinear time-domain simulations of the superradiant instability for vector fields were performed and accurate predictions for the timescale were derived~\cite{East:2017ovw,East:2017mrj}. 
The results of Refs.~\cite{East:2017ovw,East:2017mrj} are obtained in the time-domain and depend on the initial data. In particular, this means that
(i) small values of the coupling parameter $M\mu$, which lead to very long instability timescales, are not accessible and that
(ii) it is, in principle, possible that different initial conditions excite modes with an even smaller timescale.

With this is mind, the purpose of this work is two-fold: on the one hand, we close an important gap by computing the spectrum of unstable Proca modes in the frequency domain for generic values of the BH spin and of the coupling parameter $\mu M$. This allows us to estimate precisely the instability time scale for massive vector (Proca) fields around a Kerr BH in the entire parameter space of interest.
On the other hand, we use these results together with recent electromagnetic observations of accreting BHs to refine the constraints on the mass of ULVs and ALPs coming from the Regge plane.
In particular, we use the observed stability of the inner accretion disc of X-ray sources Cygnus X-1 and LMC X-3 over $14\,{\rm yr}$~\cite{Gou:2009ks} and $26\,{\rm yr}$~\cite{Steiner:2010kd}, respectively, to derive more robust constraints on the basis that --~were these BHs superradiantly unstable~-- their spin (and hence their inner accretion disc) would not be stable over this baseline.
We also discuss the potential of electromagnetic spin measurements of supermassive BHs and the projected constraints expected from the future gravitational-wave space detector LISA~\cite{Audley:2017drz}, thus extending some of the results of Ref.~\cite{Brito:2017zvb} to the vector case.

\section{Bosonic condensates around spinning BHs}

\subsection{Setup}

We focus on the following unified Lagrangian which collectively describes ALPs and dark photons (minimally) coupled to gravity,

\begin{eqnarray}
 {\cal L} &=& \sqrt{-g}\left(\frac{R}{16\pi}-\frac{1}{4}F_{\mu\nu}F^{\mu\nu}-\frac{1}{4}B_{\mu\nu}B^{\mu\nu}-\frac{1}{2}(\partial_\mu \Phi\partial^\mu \Phi+ \mu^2_{\rm a}\Phi^2)+\frac{g_{{\rm a}\gamma\gamma}}{4}\,\Phi\, F_{\mu\nu}{}^*F^{\mu\nu}\right.\nonumber\\
&&\left.+\frac{\gamma}{2} F_{\mu\nu} B^{\mu\nu}-\frac{1}{2}\mu_{\gamma}^2 B_\mu B^{\mu}\right)\,,
\end{eqnarray}
where $R$ is the Ricci curvature, $\Phi$, $A_\mu$ and $B_\mu$ are the axion-like field, the visible (Maxwell) gauge field and a hidden $U(1)$ gauge field, respectively~\cite{Goodsell:2009xc}, $F_{\mu\nu}$ and $B_{\mu\nu}$ are the corresponding field strengths. 
For definiteness, we focus on the simplest model with diagonalized mass matrix and with the simplest possible kinetic coupling matrix, which can be also obtained through a rotation in the vector field space~\cite{Goodsell:2009xc}.
The constant $g_{a\gamma\gamma}$ is the axion-photon coupling. 
In our units, the mass of the boson fields is related to the mass parameter $\mu_{a,\gamma}$ through
\be
m_{{\rm ALP},V}=\mu_{a,\gamma} \hbar\,,
\ee
where the subscript refers to ALPs and ULVs, respectively.
Astrophysical BHs are electrically neutral due to quantum discharge effects, electron-positron pair production, and charge neutralization by astrophysical plasma, and the magnetic fields of accretion disks is too weak to affect the dynamics of the system~\cite{Barausse:2014tra,Cardoso:2016olt}. 
We will therefore neglect any background electromagnetic fields; in view of the small couplings to the hidden vector, this approximation is not likely to have any impact on the physics of superradiance explored here. In this case, the field equations for the ALP and the dark photon read
\begin{eqnarray}
(\square -\mu_{\rm a}^2) \Phi&=&0  \,,  \label{eqALP}\\
\nabla_\mu B^{\mu\nu}- \mu_{\gamma}^2 B^\nu &=&0   \label{ProcaEOM}\,,
%
%
\end{eqnarray}
so they reduce to the massive Klein-Gordon and Proca equations on a curved spacetime.

Several properties of bosonic condensates near spinning BHs can be understood through a perturbative analysis. We therefore neglect the stress-energy tensors of the massive bosons which source Einstein's equations, and consider Eqs.~\eqref{eqALP}--\eqref{ProcaEOM} propagating on a Kerr metric. This approximation is consistent as long as the energy density of these fields is much smaller than that of the BH, as expected if the instability arises from a small seed (e.g., from quantum fluctuations)~\cite{Brito:2015oca}.

The Kerr BH is described by the gravitational field
\begin{equation}\label{metric}
ds^2=-\frac{\Delta}{\Sigma}\left[\mathrm{d}t-a\sin^2\theta\,\mathrm{d}\phi \right]^2+\frac{\sin^2\theta}{\Sigma}\left[(r^2+a^2)\mathrm{d}\phi -a\,\mathrm{d}t\right]^2 
+\frac{\Sigma}{\Delta}\,\mathrm{d}r^2+\Sigma\,\mathrm{d}\theta^2\,,
\end{equation}
where
\begin{equation} \label{metricAux}
\Delta=r^2+a^2-2M r\,,\qquad\qquad \Sigma=r^2+a^2 \cos^2\theta\,.
\end{equation}
The event horizon is a null hypersurface with $r=r_+$, with $r_+$ being the largest positive real root of $\Delta$. Furthermore, $M=\frac{r_+^2+a^2}{2r_+}$ is the ADM mass of the BH, $J=M a$ is its ADM angular momentum, and $|a|\leq M$, with equality saturating at extremality, where the black hole event horizon becomes degenerate. The temperature and angular velocity of the Kerr solution are 
\begin{equation} \label{KerrTOm}
T_H=\frac{r_+^2-a^2}{4\pi r_+(r_+^2+a^2)}\,,\qquad\qquad \Omega_H=\frac{a}{r^2+a^2}\,.
\end{equation}
%
%
\subsection{The spectrum of dark-photon condensates}
%
\begin{figure}[ht]
\begin{center}
\includegraphics[width=0.47\textwidth]{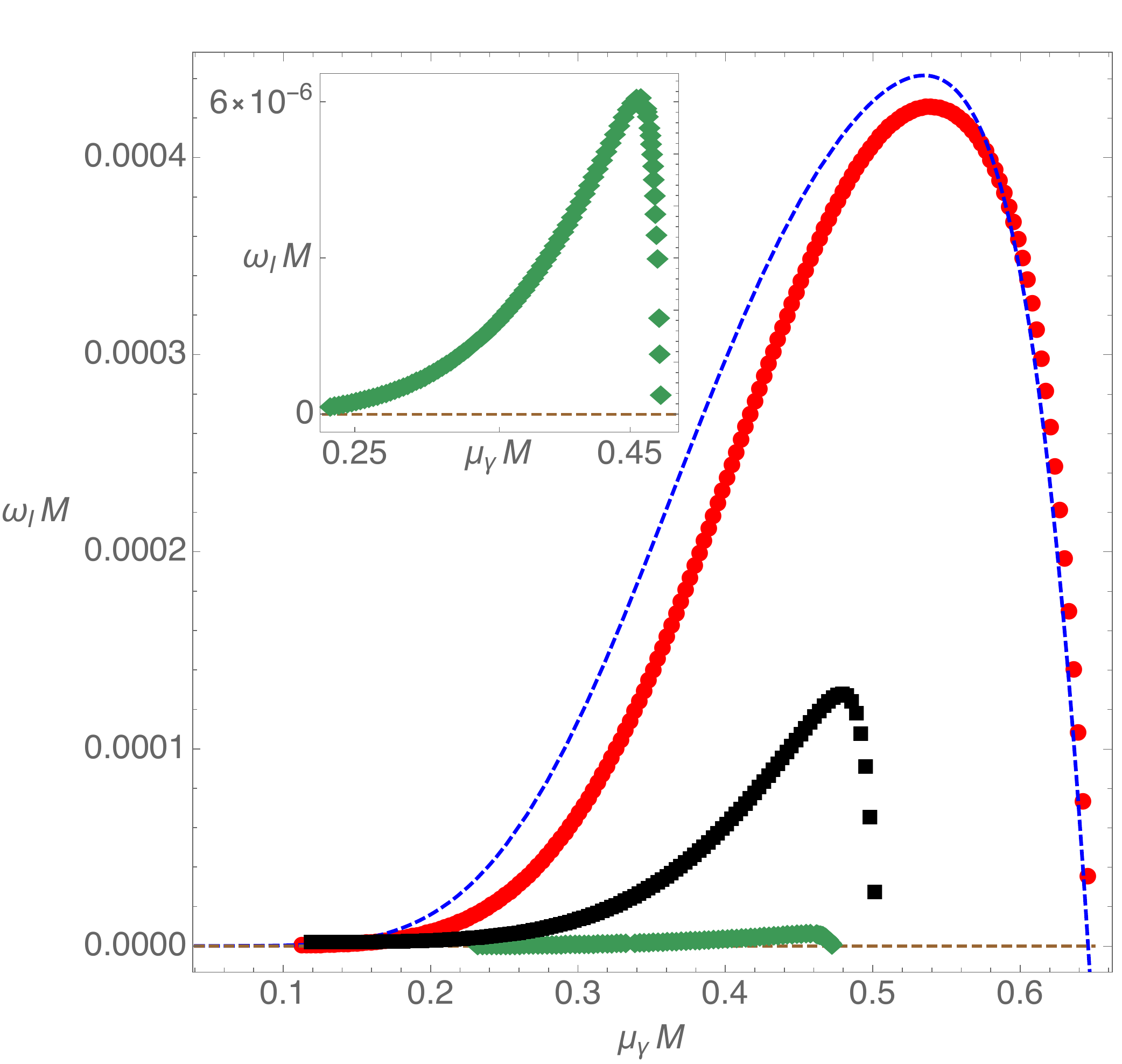}
\hspace{0.3cm}
\includegraphics[width=0.46\textwidth]{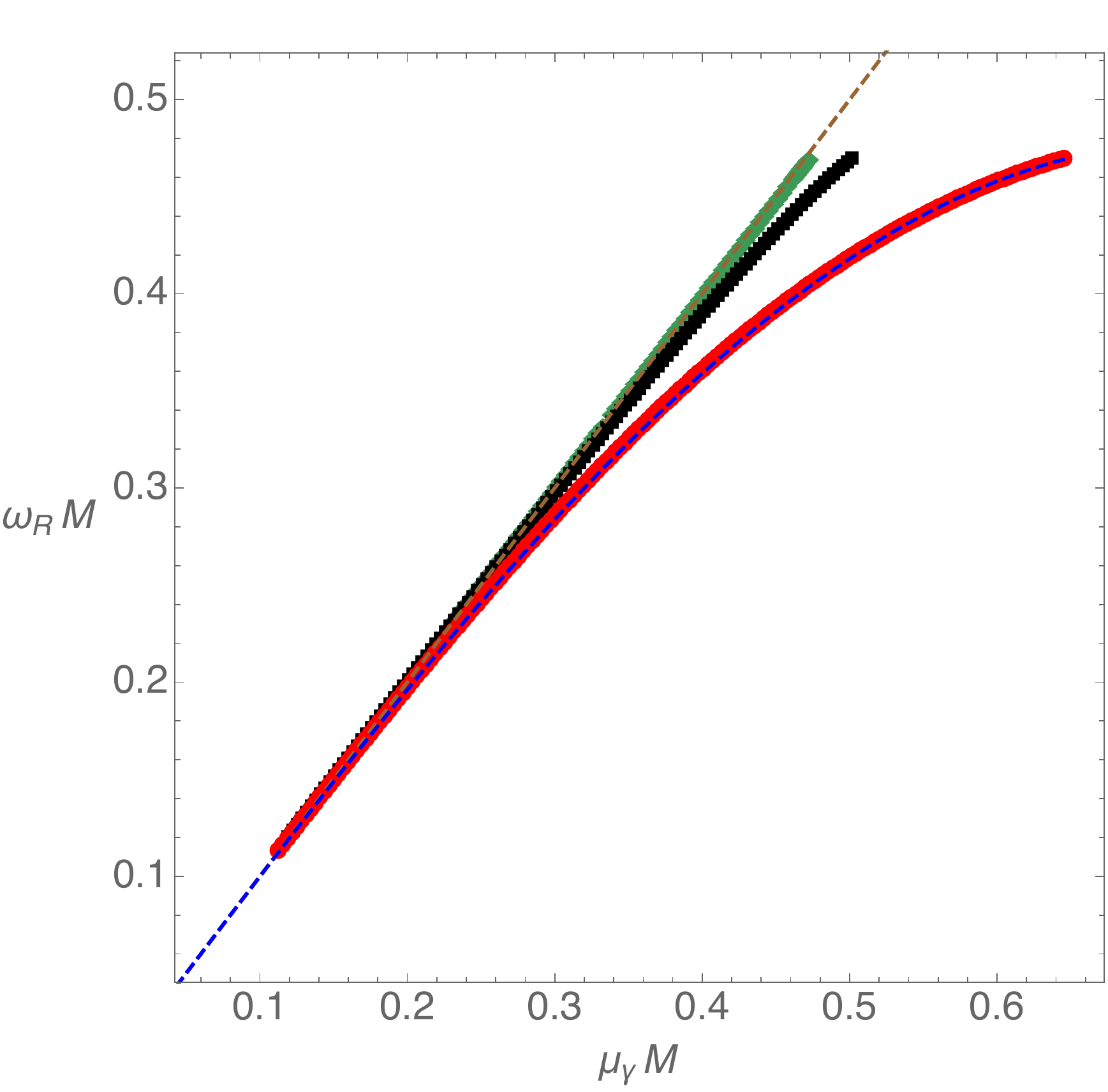}
\caption{The three sectors of massive Proca unstable modes for $a/M=0.998$ and  $m=1$.
{\it Left panel}: Instability rate $M\omega_I$ as a function of the dimensionless Proca field mass $M\mu_{\gamma}$. The inset plot is a zoom-in of the lower curve in the main plot. The blue dashed curve is the analytical fit \eqref{fitProcaI}.
{\it Right panel}: Characteristic frequency $M\omega$ as a function of the dimensionless Proca field mass $M\mu_{\gamma}$. The dashed brown curve is an auxiliary reference curve with $\omega_R=\mu_{\gamma}$. The blue dashed curve is the analytical fit \eqref{fitProcaR}.}
\label{fig:3ProcaModes}
\end{center}
\end{figure}
\begin{figure}[ht]
\begin{center}
\includegraphics[width=0.5\textwidth]{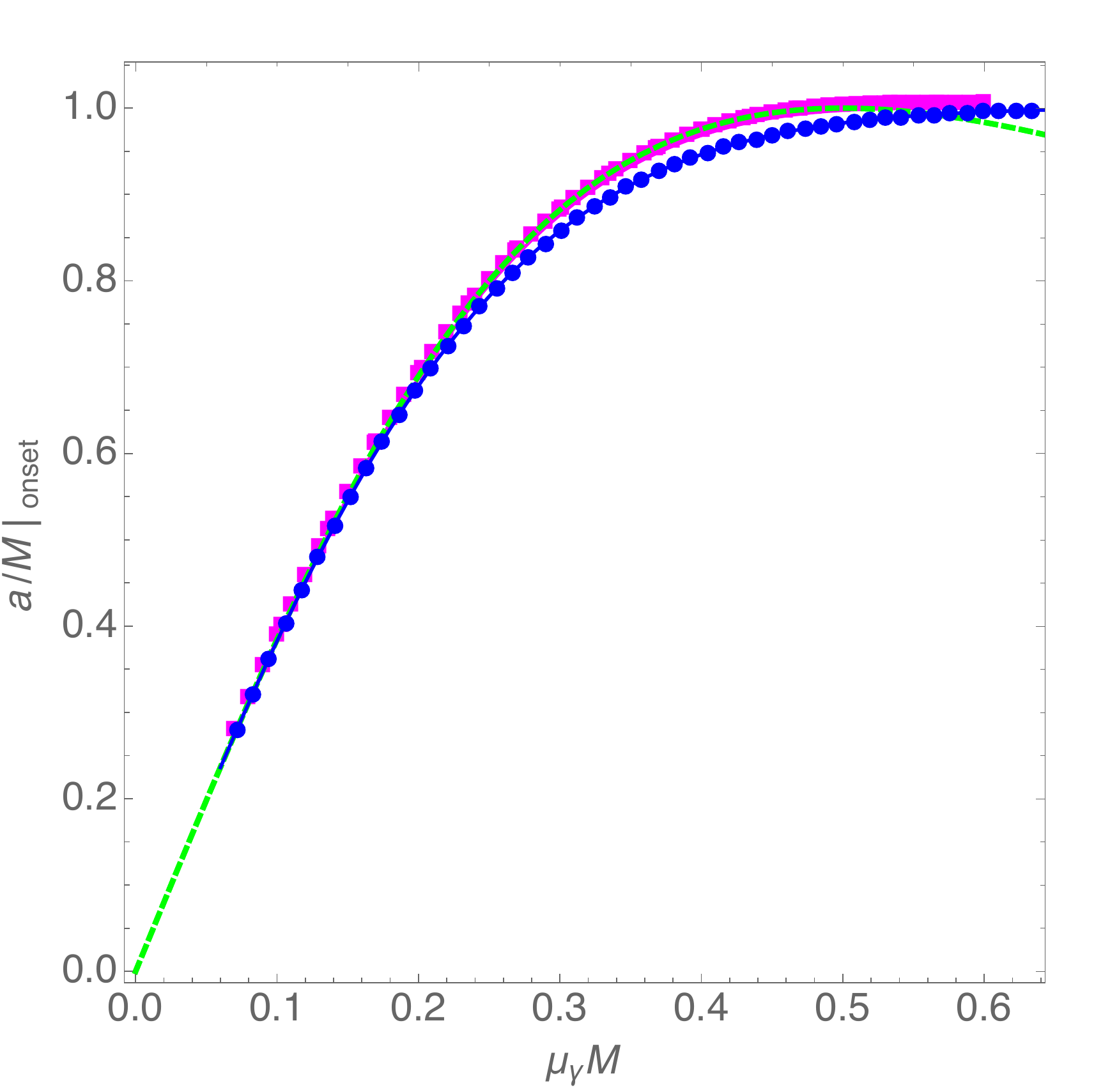}
\caption{Onset curve with $\omega_I=0$ and  $\omega_R=m\Omega_H$ and $m=1$ for the massive Proca superradiant instability for the two most unstable sectors (out of three) of perturbations. 
In particular, we focus on $S=-1,0$ and $m=1$.
The lower blue curve with disks is the onset curve for the polar $S=-1$ family of modes (see this blue curve also in Figs.~\ref{fig:ImProca} and \ref{fig:ReProca}), while the upper magenta curve with squares is the onset curve for the second most unstable family, the $S=0$ axial family (see this magenta curve also in Fig.~\ref{fig:Proca:s0}). The region above the onset curve is unstable. The disk (square) curve here  corresponds to the modes also represented with disks (squares) in Fig.~\ref{fig:3ProcaModes}. The dashed green line is the analytical curve $a/M=4\mu_\gamma M/(4\mu_\gamma^2 M^2+1)$, obtained in the small $M\mu_\gamma$ limit by requiring $\omega_R=m\Omega_H$~\cite{Pani:2012vp}.}
\label{fig:onset_Proca}
\end{center}
\end{figure}

We are interested in studying perturbations of a test Proca field with mass $\mu_{\gamma}$ on the Kerr background \eqref{metric}. The field equation for these perturbations is given in Eq.~\eqref{ProcaEOM},
with the field strength related to the potential via $B_{\mu\nu}=\partial_\mu B_{\nu}-\partial_\nu B_{\mu}$.
Since $\partial_t$ and $\partial_\phi$ are Killing vector fields of the Kerr background \eqref{metric}, one can do a Fourier decomposition of the Proca field perturbations along these directions: 
\begin{equation}\label{ProcaEOMb}
B=e^{-i\omega t}e^{i m\phi}\Big( B_t(r,\theta)\mathrm{d}t+B_r(r,\theta)\mathrm{d}r+B_\theta(r,\theta)\mathrm{d}\theta + B_\phi(r,\theta)\mathrm{d}\phi \Big),
\end{equation}
where $\omega$ and $m$ are, respectively, the frequency and azimuthal quantum number of the perturbation. 
The resulting field equations are a coupled system of four PDEs and associated boundary conditions which yield a quadratic eigenvalue problem in the frequency $\omega$, for a given mode $m$. Details of the numerical procedure \cite{Dias:2015nua,Dias:2009iu,Dias:2010maa,Dias:2010eu,Dias:2010gk,Dias:2011jg,Dias:2014eua,Dias:2010ma,Dias:2011tj,Dias:2013sdc,Cardoso:2013pza,Dias:2015wqa} that we use to solve the problem are presented in the Appendix~\ref{sec:appendix}.

\begin{figure}[th]
\begin{center}
\includegraphics[width=0.5\textwidth]{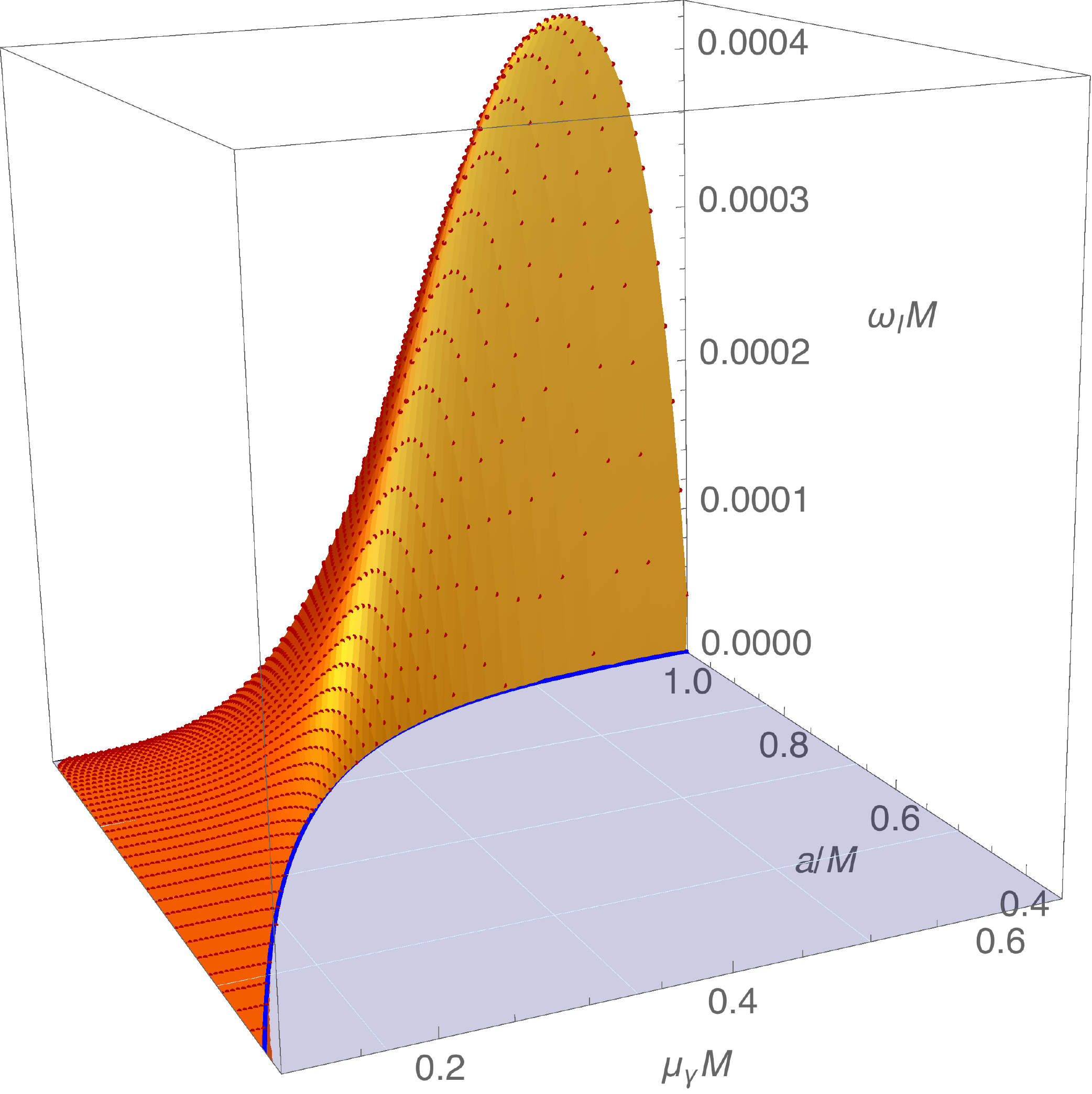}
\caption{Superradiant instability of Kerr BHs against massive vector (Proca) fields. 
The plot shows the instability rate $M\omega_I$ as a function of the dimensionless rotation $a/M$ and of the dimensionless Proca field mass $M\mu_{\gamma}$ for $-S=m=1$. The blue curve with $\omega_I=0$ signals the onset of the instability.}
\label{fig:ImProca}
\end{center}
\end{figure}
\begin{figure}[ht]
\begin{center}
\includegraphics[width=0.52\textwidth]{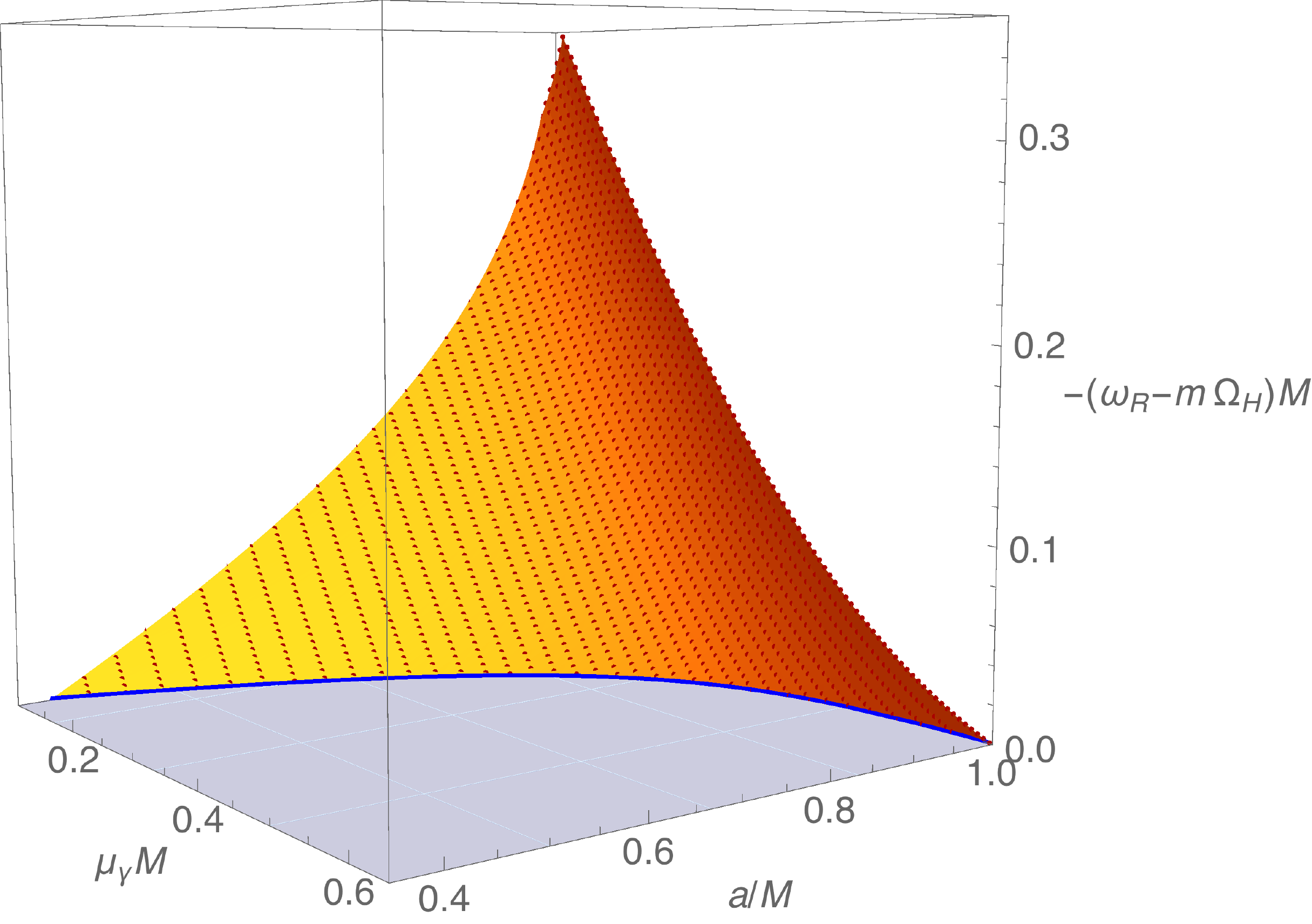}
\hspace{0.3cm}
\includegraphics[width=0.43\textwidth]{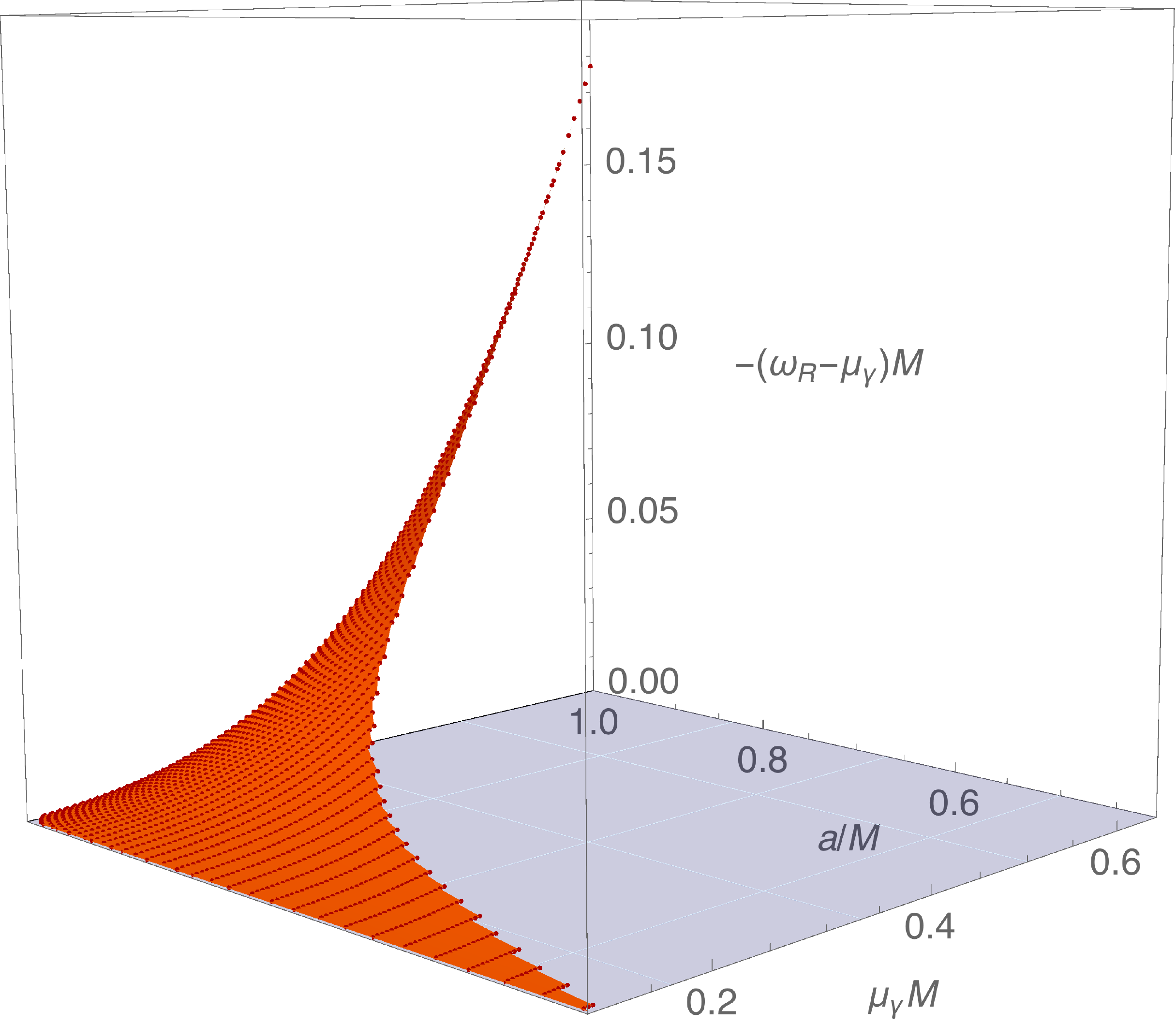}
\caption{Frequency $M\omega_R$ for the most unstable massive Proca family of modes with $-S=m=1$. 
{\it Left panel}: the difference  $\omega_R-m\Omega_H$ is always negative for the unstable modes of Fig.~\ref{fig:ImProca}. The blue curve with $\omega_R=m\Omega_H$ signals the onset of the instability already shown in Figs.~\ref{fig:onset_Proca} and \ref{fig:ImProca}.
{\it Right panel}: difference  $\omega_R-\mu_{\gamma}$ is always negative for the unstable modes of Fig.~\ref{fig:ImProca}.}
\label{fig:ReProca}
\end{center}
\end{figure}
\begin{figure}[th]
\begin{center}
\includegraphics[width=0.5\textwidth]{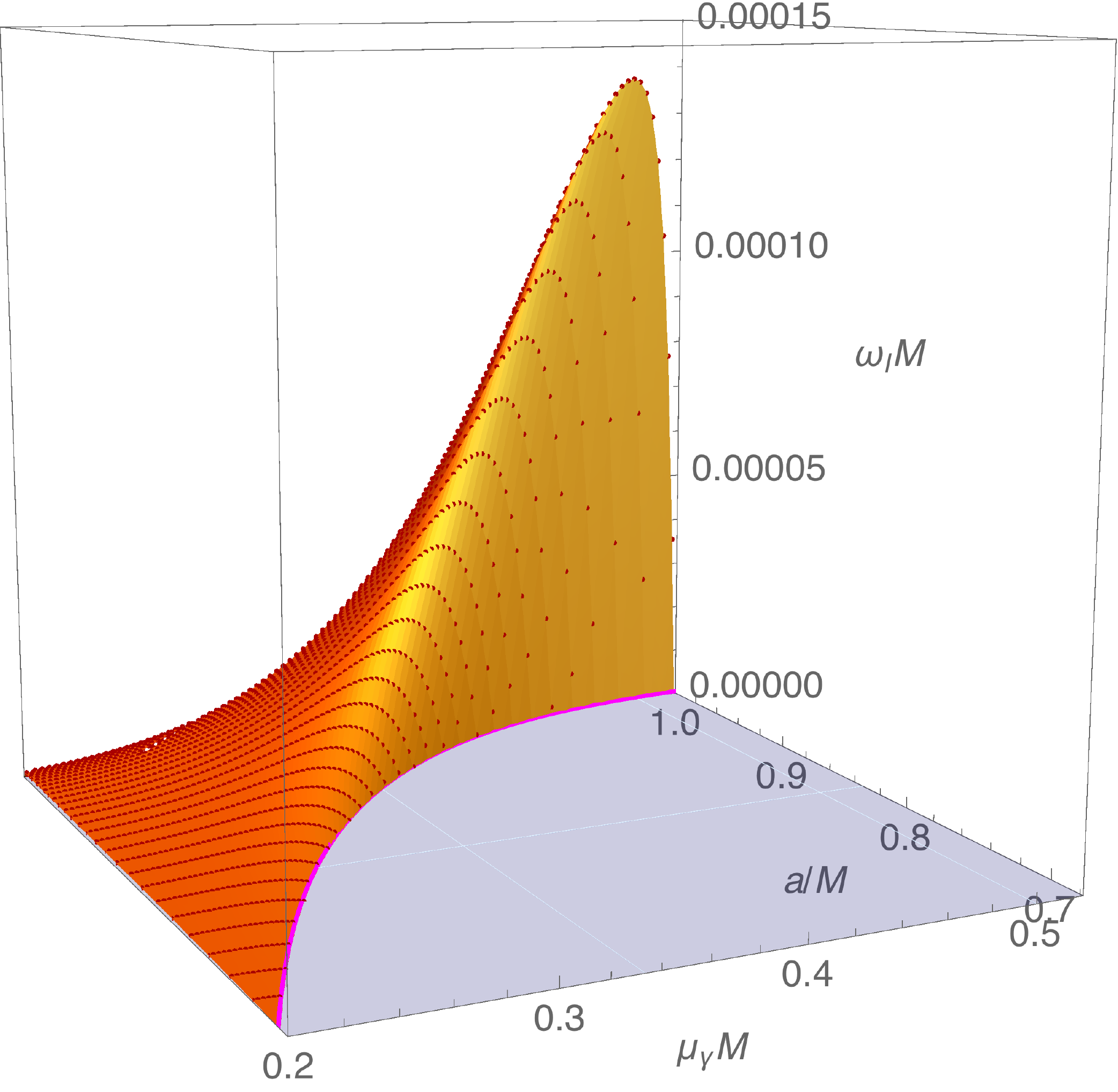}
\hspace{0.3cm}
\includegraphics[width=0.46\textwidth]{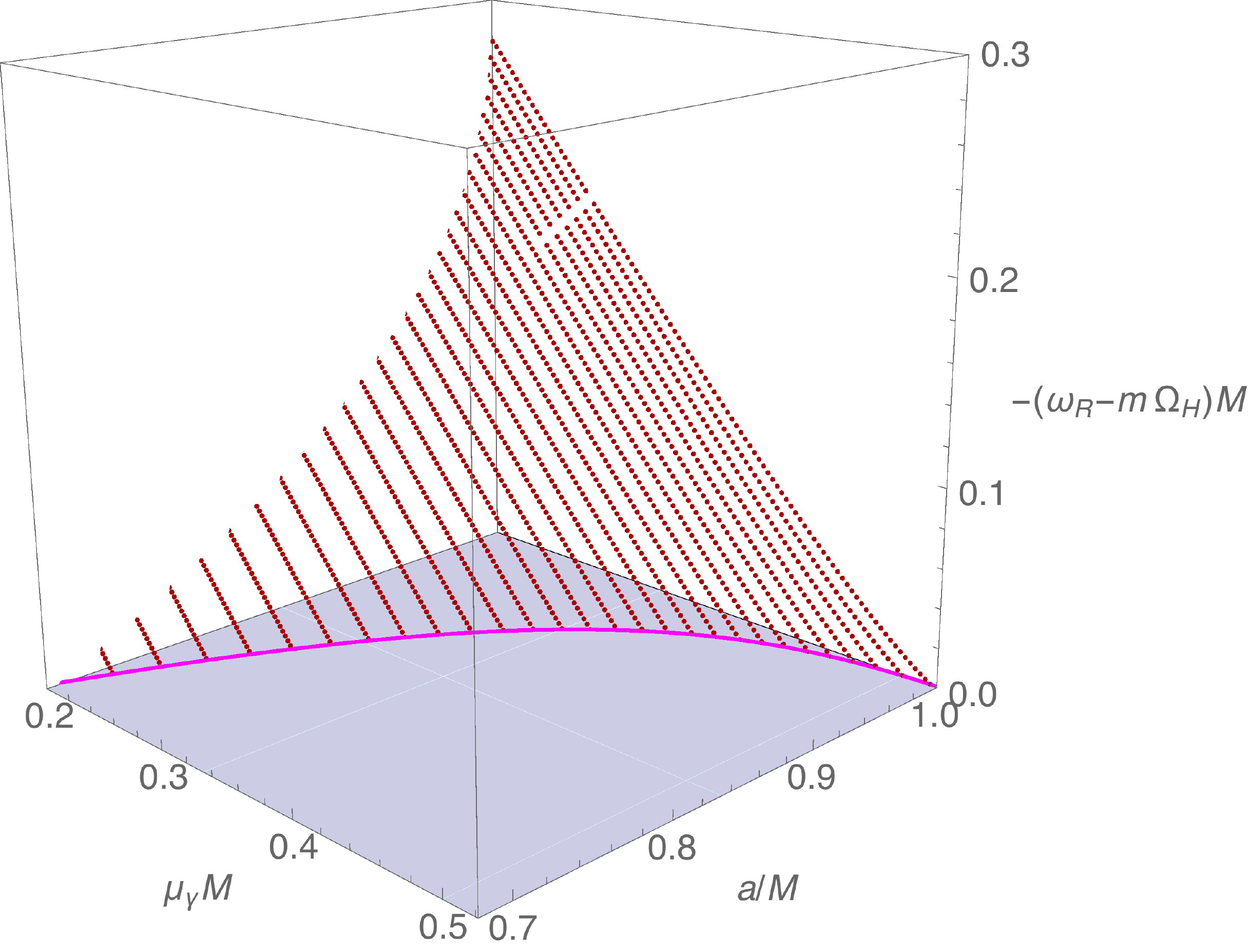}
\caption{Complex frequency $M \omega$ for the unstable massive Proca family of modes with $S=0,m=1$. 
{\it Left panel}: instability rate $M\omega_I$ as a function of the dimensionless rotation $a/M$ and of the dimensionless Proca field mass $M\mu_{\gamma}$. 
The magenta curve with $\omega_I=0$ signals the onset of the instability and was already displayed in Figs.~\ref{fig:onset_Proca}.
{\it Right panel}: the unstable modes of the left panel have $\omega_R< m\Omega_H$ (and $\omega_R<\mu_{\gamma}$).}
\label{fig:Proca:s0}
\end{center}
\end{figure}

Our numerical method allows to compute very accurately the characteristic eigenvalue frequency satisfying the appropriate boundary conditions. We write this frequency $\omega$ as
\be
\omega=\omega_R+i\,\omega_I\,.
\ee
Our main results concerning the spectra of Proca fields around Kerr BHs are summarized in Figs.~\ref{fig:3ProcaModes}-\ref{fig:Proca:s0}.
There are three families of Proca modes. To specify these three sectors it is useful to analyze their connection with the harmonic decomposition of a massive Proca field in the Schwarzschild background, i.e. in the limit where the background rotation $a$ vanishes. 
In this case the perturbations can be decomposed in a basis of spherical harmonics and we can then use the Regge-Wheeler-Zerilli \cite{Regge:1957td,Zerilli:1970se} (or the equivalent Kodama-Ishibashi \cite{Kodama:2003jz}) classification of perturbations to distinguish these modes. The three families are~\cite{Families}:

\noindent {\it i)} $S=0$ (also called axial) modes, 

\noindent {\it ii)} $S=\pm 1$ (also called polar) modes.


All the three sectors of perturbations are unstable against superradiant effects. For small Proca field masses $M\mu_\gamma$, the instability timescale is smaller for modes with azimuthal number $m=1$. Therefore we focus our attention on $m=1$ modes and discuss later the $m=2$ case\footnote{Moreover, given $m=1$, perturbative studies in the small-rotation regime \cite{Pani:2012bp,Pani:2012vp} indicate that the most unstable perturbation is the one that connects to the Schwarzschild mode $S=-1$, with harmonic index $l=1$ (or, equivalently, $\{j,\ell\}=\{1,0\}$ in the notation of Refs.~\cite{Endlich:2016jgc,Baryakhtar:2017ngi}) and overtone number $n=0$ when $a\to 0$.}.


The broad features of the superradiant instability of Proca fields are the same as scalars~\cite{Brito:2015oca}, as the reader can confirm explicitly comparing the Proca spectrum plots in Figs.~\ref{fig:ImProca}-\ref{fig:ReProca} with the scalar field ones in Figs.~\ref{fig:ImScalar}-\ref{fig:ReScalar}.
The superradiant instability is present for low-frequency modes satisfying $\omega_R<m\Omega_H$. 
For massive fields, the frequency is set by the field mass, $\omega_R\lesssim \mu_\gamma$. Thus, for very large mass couplings $M\mu_\gamma$
the superradiant condition is no longer satisfied and the instability is quenched. This is apparent in all our results.
For example, at fixed rotation the instability rate $M\omega_I$ grows with mass coupling $M\mu_\gamma$, until the threshold is saturated.
This is nicely illustrated in the left panel of Fig.~\ref{fig:3ProcaModes}, where we fix the BH rotation at $a/M=0.998$ (approximately the Thorne's limit on the spin of an accreting Kerr BH~\cite{1974ApJ...191..507T}) and we show how the instability rate, $M\omega_I$, varies as a function of the dimensionless mass coupling $M\mu_{\gamma}$ for the three families.\footnote{For each family, the most unstable modes are those with smaller number of radial zeros, i.e those that connect to the $n=0$ mode when $a\to 0$.} 
The instability rate $M\omega_I$ grows with $M\mu_\gamma$ at small mass couplings, it attains a maximum, decreases and eventually the mode becomes stable.
For this value of the rotation parameter, $M\Omega_H\sim 0.47$, thus we expect the instability to shut off at roughly $M\mu_\gamma \sim 0.5$, as confirmed by the plot. This argument can be made more precise by inspecting the behavior of the frequency $\omega_R$ as function of the mass coupling; this is shown in the right panel of Fig.~\ref{fig:3ProcaModes}: the modes do obey to good precision $\omega_R\lesssim \mu$ (but see below for more accurate quantitative fits), and modes where this relation is observed to a good precision are also the modes where the instability turns off very close to $M\mu_\gamma\sim 0.47$. To sum up, as expected for superradiant modes that are trapped by the potential barrier created by the Proca field mass, Fig.~\ref{fig:3ProcaModes} shows that one always has $\omega_R< m\Omega_H$ and $\omega_R<\mu_{\gamma}$ (cf.\ also  Fig.~\ref{fig:ReProca} below).

Figure~\ref{fig:3ProcaModes} is also interesting in another perspective: it shows that there is a clear distinction between the three different families of modes. The curves never intersect: the most unstable, $S=-1$ family (upper red curve with disks) is the same for any $\mu_{\gamma}$. Moreover, this family is also the one that is unstable up to higher values of the Proca mass $\mu_{\gamma}$. This result will be important when deriving constraints on the ULV mass (cf.\ Sec.~\ref{sec:constraints}) because it allows us to focus on the most unstable family ($S=-1$) in the entire parameter space.

In addition, this very same family is also the one that, for a given Proca mass $\mu_{\gamma}$, starts becoming unstable at a smaller value of the rotation. This is best illustrated in Fig.~\ref{fig:onset_Proca} where we plot the threshold line for the superradiant instability in a $(a/M,M\mu_\gamma)$ plane, for the two most unstable sectors (namely, the $S=-1$ and $S=0$ families represented by the disks and squares in Fig.~\ref{fig:3ProcaModes}). In other words, this is the line for which $\omega_I=0$ and $\omega_R=m\Omega_H$. Above this curve, the Kerr BH is unstable. The lower blue curve with disks (with  $S=-1$) is the most unstable family and it is also the one that becomes unstable at (slightly) lower values of $a/M$.    

The full dependence of the instability rate on $M\mu_\gamma$ and on the BH rotation $a/M$ is shown in Fig.~\ref{fig:ImProca} for the
family with the smallest instability timescale, i.e. modes with $-S=m=1$. For fixed Proca field mass $M\mu_{\gamma}$, the instability typically grows with rotation (up to a point very close to extremality where it reaches a maximum, see below), as might be anticipated. Among the particular modes that we computed {\it explicitly}, we find that the most unstable mode occurs for 
\begin{equation} \label{maxInstability}
\{\mu_{\gamma}M, a/M\} = \{0.539459, 0.998027\} \quad \hbox{with} \quad \omega M = 0.435065 +4.256752\times 10^{-4}\,i \,,
\end{equation}
which corresponds to an instability timescale
\begin{equation}
 \tau_{\rm inst}\equiv\frac{1}{\omega_I}\approx 0.1\left(\frac{M}{10 M_\odot}\right)\,{\rm s}\,.
\end{equation}


For reference, we also give the data for three other modes:
\begin{eqnarray} \label{examplesws}
&& \{\mu_{\gamma}M, a/M\} = \{0.301385, 0.991758\} \quad \hbox{with} \quad \omega M = 0.285193 +6.805245\times 10^{-5}\,i \nonumber\\
&& \{\mu_{\gamma}M, a/M\} = \{0.393974, 0.988584\} \quad \hbox{with} \quad \omega M = 0.355112 +2.246362\times 10^{-4}\,i \nonumber\\
&& \{\mu_{\gamma}M, a/M\} = \{0.498262, 0.988584\} \quad \hbox{with} \quad \omega M = 0.415059 +3.019861\times 10^{-4}\,i \,.\nonumber
\end{eqnarray}
This data is consistent with the findings of \cite{East:2017mrj} where the superradiant instability timescales were read from time domain simulations. Indeed, Ref.~\cite{East:2017mrj} finds: i) $M\omega_I \sim 7\times 10^{-5}$ for $\left(M\mu_{\gamma}, a/M\right)=(0.3,0.99)$,  ii) $M\omega_I \sim 2\times 10^{-4}$ for $\left(M\mu_{\gamma}, a/M\right) =(0.4,0.99)$ and  iii) $M\omega_I\sim 3\times 10^{-4}$ for $\left(M\mu_{\gamma}, a/M\right)=(0.5,0.99)$.

For completeness, in Fig.~\ref{fig:ReProca} we also give the real part of the frequency $M\omega_R$ of the unstable modes considered in Fig.~\ref{fig:ImProca}. We do so by comparing $M\omega_R$ with the quantity $m\Omega_H$ (left panel) and with the Proca mass $\mu_{\gamma}$ (right panel). As expected our unstable modes satisfy the superradiant condition $\omega_R< m\Omega_H$ and they have $\omega_R<\mu_{\gamma}$ which indicates that they are trapped by the potential barrier created by the Proca field mass. The onset curve (blue) of the instability has  $\omega_R=m\Omega_H$ and coincides with the onset curve (blue) of Fig.~\ref{fig:ImProca}, namely $\omega_I=0$. This onset curve is also the blue curve with disks in Fig.~\ref{fig:onset_Proca}.\footnote{For the benefit of the reader interested on using or reproducing our results, the list of unstable modes plotted in Figs.~\ref{fig:ImProca} and \ref{fig:ReProca} is available in a file named "data" with the format $\{\mu_{\gamma}M,a/M,\omega M \}$ that is included in the {\tt arXiv} version of this manuscript. Our pseudospectral method has exponential convergence and we used quadruple precision. This guarantees that our results for the frequency are accurate up to, at least, 8 decimal digits.}

Finally, for the sake of completeness, we also take the opportunity to show the full frequency spectrum of superradiant modes of the family with $S=0$ and $m=1$ in Fig.~\ref{fig:Proca:s0} (a particular curve with $a/M=0.998$ of this second most unstable family was already displayed in the black curve of Fig.~\ref{fig:3ProcaModes}). Note that the onset curve (magenta) of the instability with $\omega_R=m\Omega_H$ and $\omega_I=0$ is also the magenta onset curve with squares already shown in Fig.~\ref{fig:onset_Proca}.

\subsubsection{Fitting formulas for the most unstable Proca modes of a Kerr BH}
While the behavior of $\omega_R$ is regular and straightforward to fit, the behavior of $\omega_I$ as a function of $a/M$ and $\mu_{\gamma}M$ is more involved. 
For the most unstable family of modes with $-S=m=1$ and $n=0$, we find that the following functions
\begin{eqnarray}
M\omega_R &\simeq& M\mu_{\gamma}\left(1+\alpha_1M\mu_{\gamma}+\alpha_2(M\mu_{\gamma})^2+\alpha_3(M\mu_{\gamma})^3\right)\,, \label{fitProcaR}\\
M\omega_I &\simeq& \beta_0\, (M\mu_{\gamma})^7\left(1+\beta_1M\mu_{\gamma}+\beta_2(M\mu_{\gamma})^2\right) \left(\chi- 2 \omega_R r_+\right) \,, \label{fitProcaI}
\end{eqnarray}
provide a good fit of our numerical results, as shown in Fig.~\ref{fig:3ProcaModes}. In the above equations, the coefficients $\alpha_i(\chi)$ and $\beta_i(\chi)$ are functions of the dimensionless spin $\chi\equiv a/M$. For these coefficients we find the following polynomial fit in powers of $\sqrt{1-\chi^2}$,
\begin{eqnarray}
 \alpha_i &=& \sum_{j=0}^4 A_j^{(i)}(1-\chi^2)^{j/2}\,,\qquad \beta_i =\sum_{j=1}^4 B_j^{(i)}(1-\chi^2)^{j/2}+\sum_{j=0}^4 C_j^{(i)} \chi^{j}\,, \label{alphabetafit}
\end{eqnarray}
where the coefficients $A_j^{(i)}$, $B_j^{(i)}$ and $C_j^{(i)}$ are given in Table~\ref{tab:fit} in Appendix~\ref{app:fit}, where we provide more details on the fit. 
As discussed in Appendix~\ref{app:fit}, in a large region of the parameter space the precision of the fit is at least $0.2\%$ ($50$\%) for the real (imaginary) part of the frequency. 

%
Unfortunately, we are unable to extract accurate results for coupling $M\mu_\gamma \lesssim 0.06$, so the above fit should be extrapolated with some care when $\mu_\gamma M\ll0.1$. Nonetheless, it agrees with known analytical results when $\mu_\gamma M\ll1$, as discussed in the next section.


\subsubsection{Comparison to previous results}
Previous studies in the frequency domain either approximated the background as nearly Newtonian $(M\mu_\gamma\ll1$) or through a slow-rotation expansion ($a/M\ll1$).
In the small-coupling limit, studies of Proca field around slowly spinning Kerr BHs suggested that~\cite{Pani:2012vp,Pani:2012bp}
\be
M\omega_I\simeq 2\gamma_{Sm}r_+\left(m \Omega_H-\omega_R \right) (M\mu_\gamma)^{4m+5+2S}\,,\label{genericb}
\ee
where $S$ labels the family. The dependence on the family was further established via perturbative studies around nearly Newtonian backgrounds~\cite{Endlich:2016jgc,Baryakhtar:2017ngi}.
The results in the literature for $\gamma_{Sm}$ are summarized in Table~\ref{ringtable1}.
\begin{table}
\centering \caption{Coefficient $\gamma_{Sm}$ of \eqref{genericb} as computed with different methods or approximations.
The coefficient $\beta_0$ depends on the spin $\chi\equiv a/M$ and roughly ranges between $2$ and $6$ for $\chi\in(0,1)$.
A question mark indicates that the results has not been computed or is not enough accurate within the corresponding approximation scheme. 
} 
\vskip 12pt
\begin{tabular}{@{}c|c|ccc@{}}
\hline \hline
Reference& Approximation &\multicolumn{3}{c}{$\gamma_{S1}$}\\ \hline
						  &      		&$S=-1$             & $S=0$     & $S=1$ \\
Pani {\it et al.}~\cite{Pani:2012vp,Pani:2012bp}  & $a/M\ll1$, numerical to second order           &$20\pm 10$         &$1/12$     &?        \\
Endlich and Penco~\cite{Endlich:2016jgc}          & $a/M\ll1$, analytical to first order		&$20/3$             &$1/3$      &$320/59049$  \\
Baryakhtar {\it et al.}~\cite{Baryakhtar:2017ngi} & $M\mu_\gamma\ll1$, analytical to leading order             	&$4$               &$1/6$       &$?$        \\
This work                                         & $M\mu_\gamma\gtrsim 0.06$, exact   			&$\beta_0(\chi)$          &?        &$?$        \\
\hline \hline
\end{tabular}
\label{ringtable1}
\end{table}  
Our results, with the provisos above, yield $\gamma_{-11}=\beta_0(\chi)$, which roughly ranges between $2$ and $6$ for $\chi\in(0,1)$ (cf.\ Eq.~\eqref{alphabetafit} and Table~\ref{tab:fit}). Although our data are in fact complementary to the $M\mu_\gamma\ll1$ regime, it is reassuring that an extrapolation to $M\mu_\gamma\to0$ is consistent with existing analytical results.
In Sec.~\ref{sec:constraints}, we will use the unstable modes to derive constraints on the ULV mass using BH mass and spin measurements. 


Finally, we have also tried to fit the modes for $S=0$ using the fitting function~\eqref{genericb}. In this case, the fit is less accurate and it is not listed in Table~\ref{ringtable1}. We suspect that the $S=0$ Proca instability has a behavior similar to the ultralight-scalar instability (reviewed in the next subsection). In particular, $\omega_I$ might display a more complex behavior as a function of the spin for moderately large values of $a/M$, as suggested by Eq.~\eqref{wI} below for the scalar case. Nonetheless, as previously shown, the $S=0$ Proca modes are always subleading relative to the $S=-1$ modes, so they play a negligible role for the bounds discussed in Sec.~\ref{sec:constraints}.

\subsection{The spectrum of ALP condensates}\label{sec:spectrumscalar}
We summarize here the known properties of the instability for ultralight scalars. In this case, the solutions are known well, both numerically and analytically.
The scalar can be expanded in scalar spheroidal harmonics labeled by the ``usual'' angular numbers $l$ and $m$. The solution for the characteristic frequencies is labeled with one more
integer, the overtone number $n$.

\begin{figure}[ht]
\begin{center}
\includegraphics[width=0.5\textwidth]{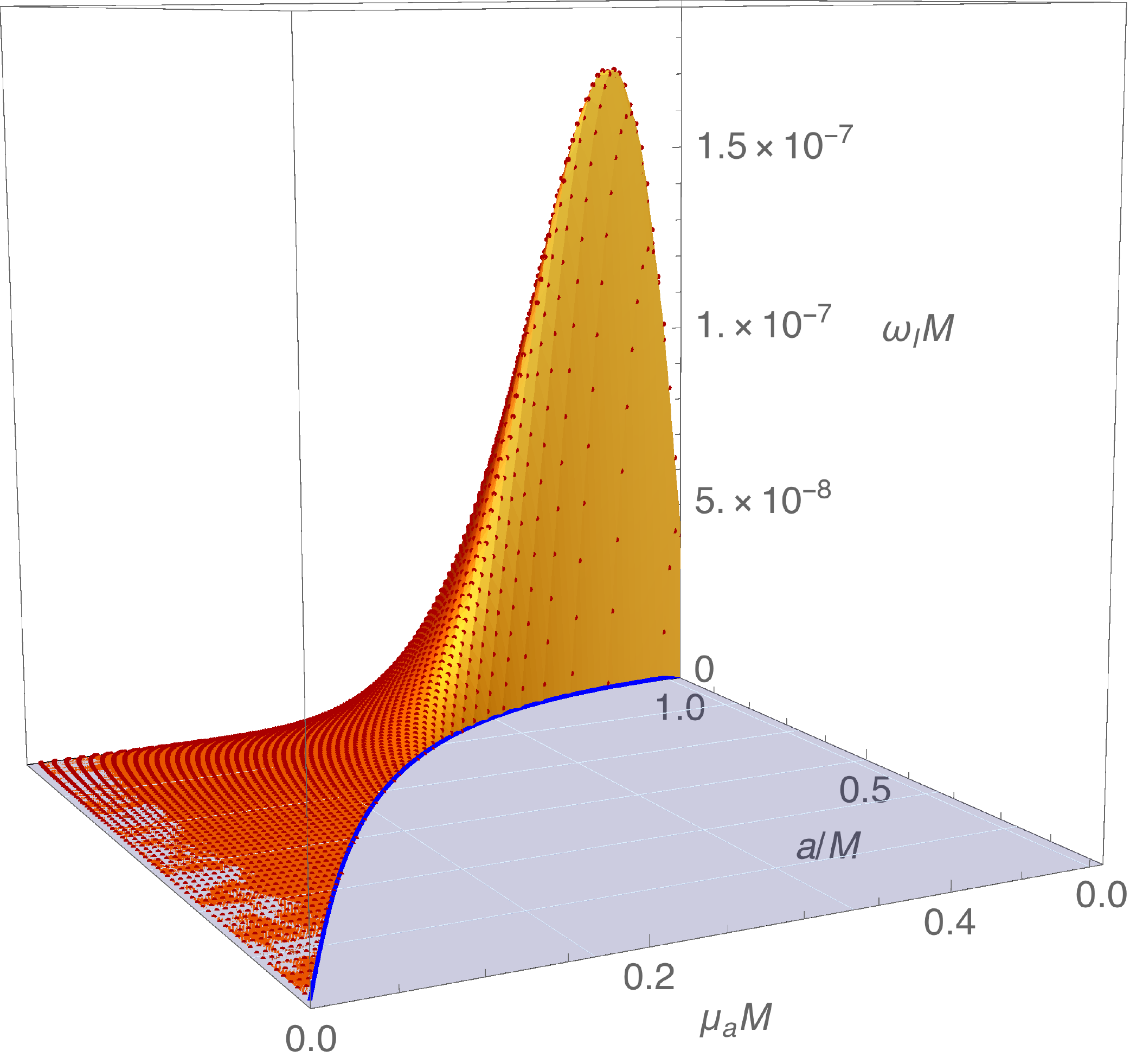}
\caption{Superradiant instability of Kerr BHs against massive scalar fields. 
The plot shows the instability rate $M\omega_I$ as a function of the dimensionless rotation $a/M$ and of the dimensionless scalar field mass $\mu M$ for $\{l,m,n\}=\{1,1,0\}$. The blue curve with $\omega_I=0$ signals the onset of the instability.}
\label{fig:ImScalar}
\end{center}
\end{figure}

\begin{figure}[hb]
\begin{center}
\includegraphics[width=0.52\textwidth]{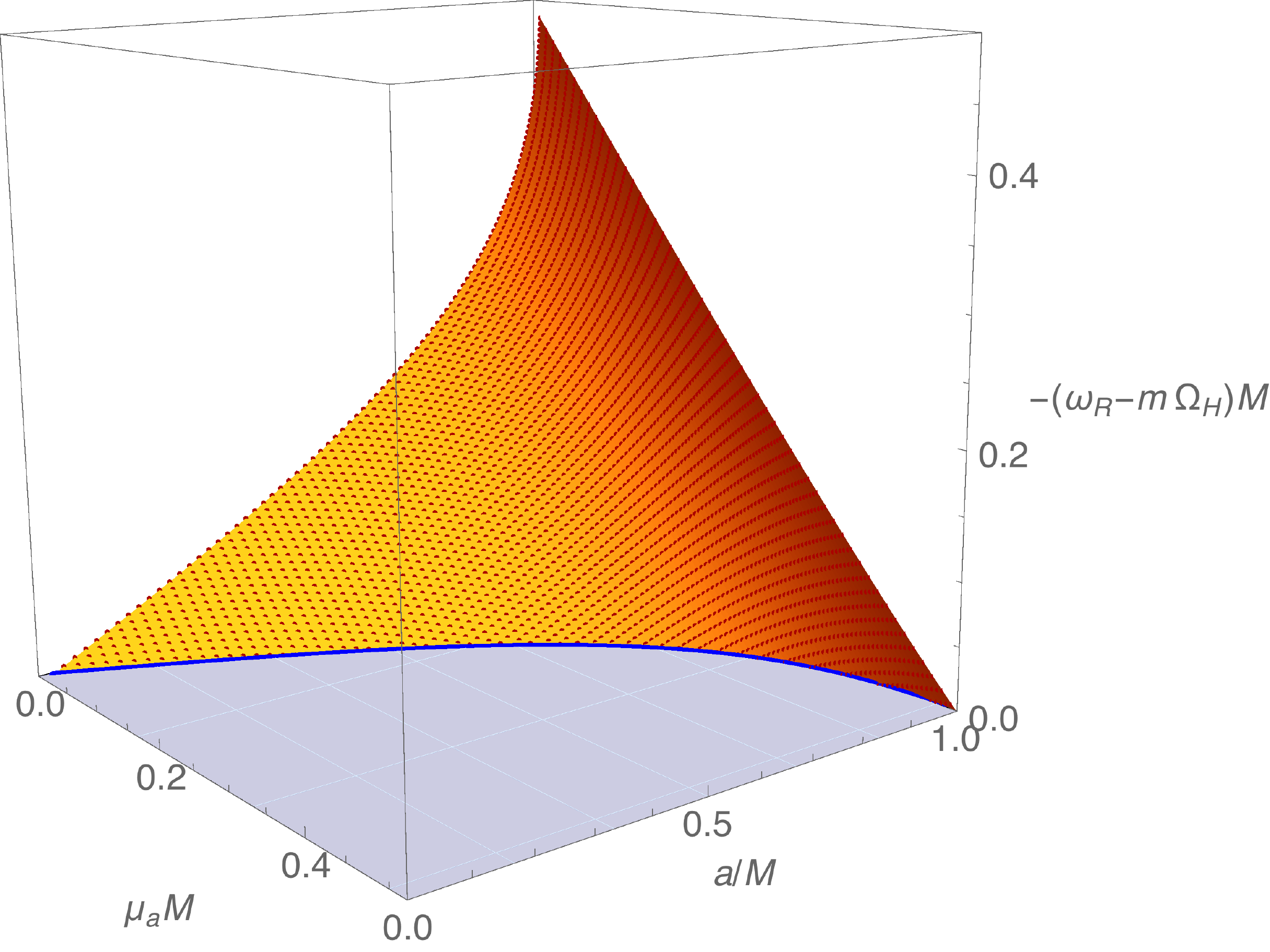}
\hspace{0.3cm}
\includegraphics[width=0.43\textwidth]{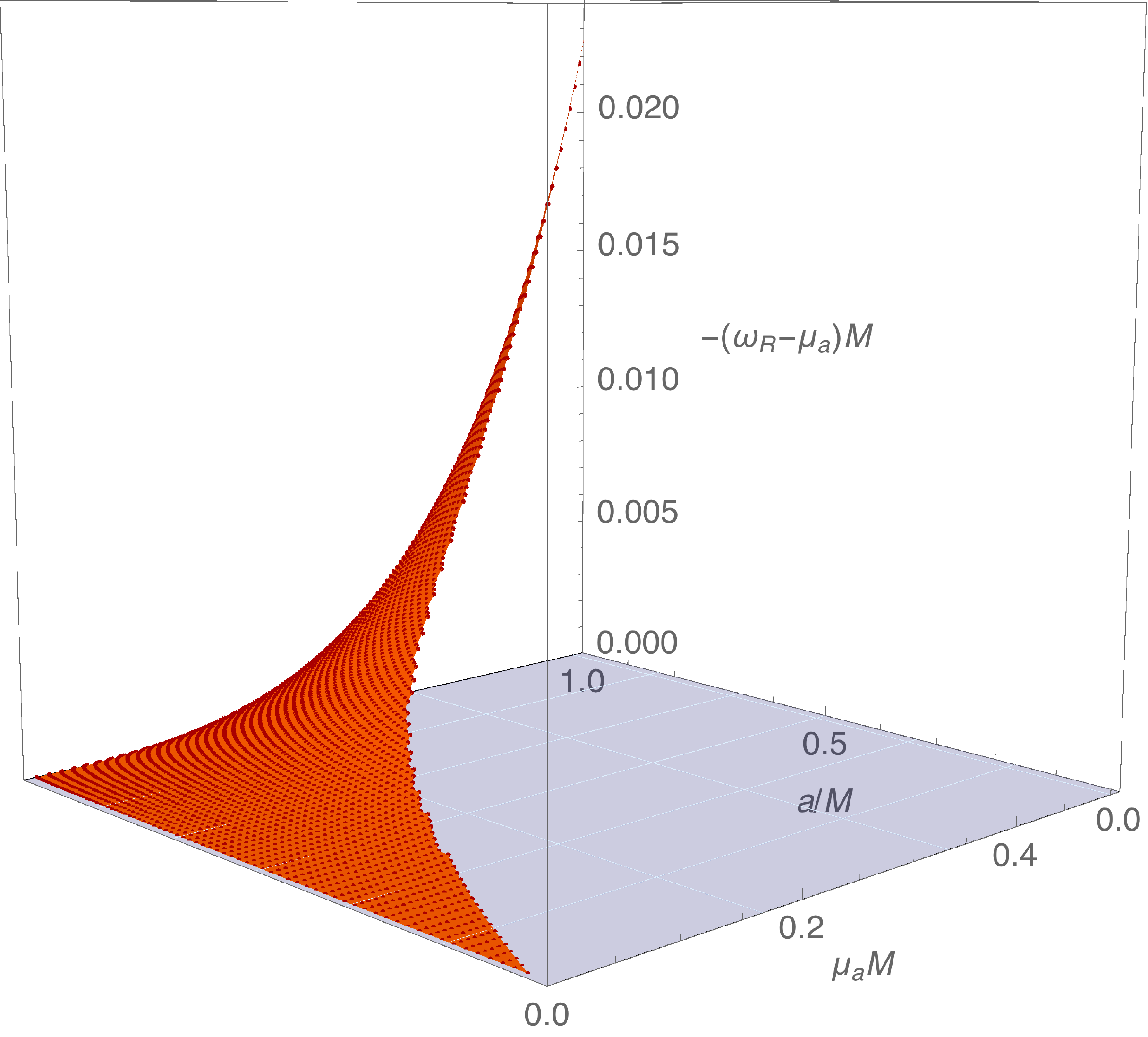}
\caption{Frequency $M\omega_R$ for unstable massive scalar field modes with $\{l,m,n\}=\{1,1,0\}$. 
{\it Left panel}: the difference  $\omega_R-m\Omega_H$ is always negative for the unstable modes of Fig.~\ref{fig:ImScalar}. The blue curve with $\omega_R=m\Omega_H$ signals the onset of the instability already shown in Fig.~\ref{fig:ImProca}.
{\it Right panel}: difference  $\omega_R-\mu_{a}$ is always negative for the unstable modes of Fig.~\ref{fig:ImScalar}.}
\label{fig:ReScalar}
\end{center}
\end{figure}

Nevertheless, for completeness, we take the opportunity to present the full spectrum of characteristic frequencies for the the most unstable scalar field mode, namely with $\{l,m,n\}=\{1,1,0\}$ (as far as we are aware this spectrum was never shown for the full phase space  $\{\mu_a,a\}$). The imaginary part of the frequency is given in  Fig.~\ref{fig:ImScalar}, while the real part of the frequency is displayed in   Fig.~\ref{fig:ReScalar}. Among the particular modes that we have computed {\it explicitly}, we find that the most unstable mode occurs for 
\begin{equation} \label{maxInstability2}
\{\mu_a M, a/M\} = \{0.448262, 0.995656\} \quad \hbox{with} \quad \omega M = 0.474649 +1.704340\times 10^{-7}\,i \,,
\end{equation}
and its associated instability timescale is roughly $2500$ times longer than in the vector case.

%


The characteristic frequencies can also be computed analytically in a perturbative expansion about the scalar field mass. In the nonrelativistic regime, $M\mu_a\ll l$, the eigenfrequencies read~\cite{Detweiler:1980uk,Brito:2015oca,Pani:2012bp}
\begin{eqnarray}
 \omega_R &\sim& \mu_a\left(1-\frac{(M\mu_a)^2}{2(l+1+n)^2}\right) \,, \label{wR} \\
M\omega_I &\sim& D_{lmn}r_+ (m\Omega_H-\omega_R) (M\mu_a)^{4l+5}\,, \label{wI}
\end{eqnarray}
where 
\begin{equation}
 D_{lmn} = \frac{2^{4l+2}(2l+n+1)!}{(l+1+n)^{2l+4}n!}\left[\frac{l!}{(2l)!(2l+1!)}\right]^2 \prod_{j=1}^l\left[j^2\left(1-a^2/M^2 \right)+4 r_+^2(\omega_R-m\Omega_H)^2\right]\,.
\end{equation}
In the following we will focus on fundamental ($n=0$) modes with $m=l$ since these are the modes with the shorter instability time scale\footnote{Near extremality, higher modes with $n>0$ might have slightly shorter time scale than $n=0$ modes~\cite{Yoshino:2013ofa,Yoshino:2015nsa}. Because the order of magnitude of the time scale is the same, we ignore here this possibility.}.

As clear from Eq.~\eqref{wI}, the time scale associated with higher-$l$ modes is considerably longer than that associated with the dominant $l=1$ modes, namely
\begin{equation}
 \frac{\tau_{{\rm inst},l+1}}{\tau_{{\rm inst},l}}\sim (M\mu_a)^4\,.
\end{equation}
Therefore, at least in the nonrelativistic regime $M\mu_a\ll 1$, modes with different harmonic index $l$ have very different time scales and can be treated separately. This separation of time scales is also confirmed by an exact numerical analysis~\cite{Yoshino:2013ofa,Brito:2015oca,Yoshino:2015nsa}.  
On the other hand, for $l=m$ the unstable modes must satisfy $\omega_R< l\Omega$, so there might be regions of the parameter space in which $l=m=1$ modes are stable but modes with higher $l=m$ are not~\cite{Dolan:2007mj}. 

\section{Constraining ultralight bosons with BH spin measurements} \label{sec:constraints}
%
A generic prediction of the BH superradiant instability is the fact that --~in the presence of ultralight bosons~-- highly spinning BHs would lose angular momentum through the instability over a timescale that might be much shorter than typical astrophysical timescales --~although parametrically longer than the dynamical timescale of the BH, the latter being ${\cal O}(M)$.
Thus, accurate measurements of the mass and spin of astrophysical BHs can be turned into indirect constraints on the mass of ultralight bosons~\cite{Arvanitaki:2010sy}.

BH spin is routinely obtained from the electromagnetic spectrum using reliable proxies for the position of the innermost stable circular orbit (ISCO)~\cite{Middleton:2014sma}. Both the (mass-independent) shape of the iron K$_{\alpha}$ line seen in reflection~\cite{Fabian:2000nu} and the thermal emission from the inner edge of the disc~\cite{Zhang:1997dy} --~assumed to be the ISCO~\cite{Shafee:2007sa,Penna:2010hu} (see also \cite{Zhu:2012vf} for a discussion on the emission from within the plunging region)~-- are commonly employed for both stellar mass BHs in binaries and supermassive BHs in active galactic nuclei (although the latter's use is a more recent development~\cite{Done:2011at}). Whilst important caveats exist for both traditional approaches, convincing evidence for truncation at the ISCO comes from the consistent position of the inner disc edge in LMC X-3 from modeling of the thermal dominant state, providing a remarkably stable spin value over a baseline of $26~{\rm yr}$~\cite{Steiner:2010kd}. Whilst not as well sampled (due to the source rarely entering the requisite thermal dominant state), Cyg X-1 also shows a remarkably stable spin value over $14~{\rm yr}$ from the same approach~\cite{Gou:2009ks}.
These observations imply that at the moment LMC X-3 and Cyg X-1 are not undergoing a superradiant instability \emph{at least} over a timescale of $26~{\rm yr}$ and $14~{\rm yr}$, respectively. Below, we will use this fact --~together with our computation of the instability timescale done in the previous section~-- to put \emph{direct} constraints on the mass of ULVs and ALPs.

More stringent (albeit less direct) constraints come from comparing the instability timescale against a typical accretion timescale, that we estimate here to be the Salpeter time, 
\begin{equation}
t_S=4.5\times 10^8 {\rm\, yr\,} \frac{\eta}{f_{\rm Edd} (1-\eta)}\,, \label{salpeter}
\end{equation}
where $f_{\rm Edd}$ is the Eddington ratio for mass accretion, and the
thin-disk radiative efficiency $\eta\equiv1-E_{_{\rm ISCO}}$ is a function of the spin related to the specific
energy $E_{_{\rm ISCO}}$ at the ISCO~\cite{Bardeen:1972fi}.

A novel approach to measure the masses and spins of astrophysical BHs come from gravitational-wave astronomy. Binary BHs are arguably the cleanest gravitational sources so measurements of the mass and spin of the binary components should be less affected by systematics than in the electromagnetic case.
Whilst the spins of the primary and secondary objects in the coalescence events detected so far by LIGO are affected by large uncertainties and are anyway (marginally) compatible to zero spin, future detections will provide more stringent constraints on the individual spins, at the level of $30\%$~\cite{TheLIGOScientific:2016pea}.
More precise measurement will come from the LISA space mission~\cite{Audley:2017drz}. LISA will be able to measure the mass and spin of binary BH components out to cosmological distances. Depending on the mass of BH seeds in the early Universe, LISA will also detect intermediate mass BHs, thus probing the existence of
light bosonic particles in a large mass range (roughly $m_s\sim 10^{-13}$--$10^{-16}$~eV) that is inaccessible to electromagnetic observations of stellar and
supermassive BHs and to Earth-based gravitational-wave detectors~\cite{Brito:2017wnc,Brito:2017zvb}.
\subsection{Constraints on dark photons}
%
\begin{figure}[th]
\begin{center}
\includegraphics[width=0.47\textwidth]{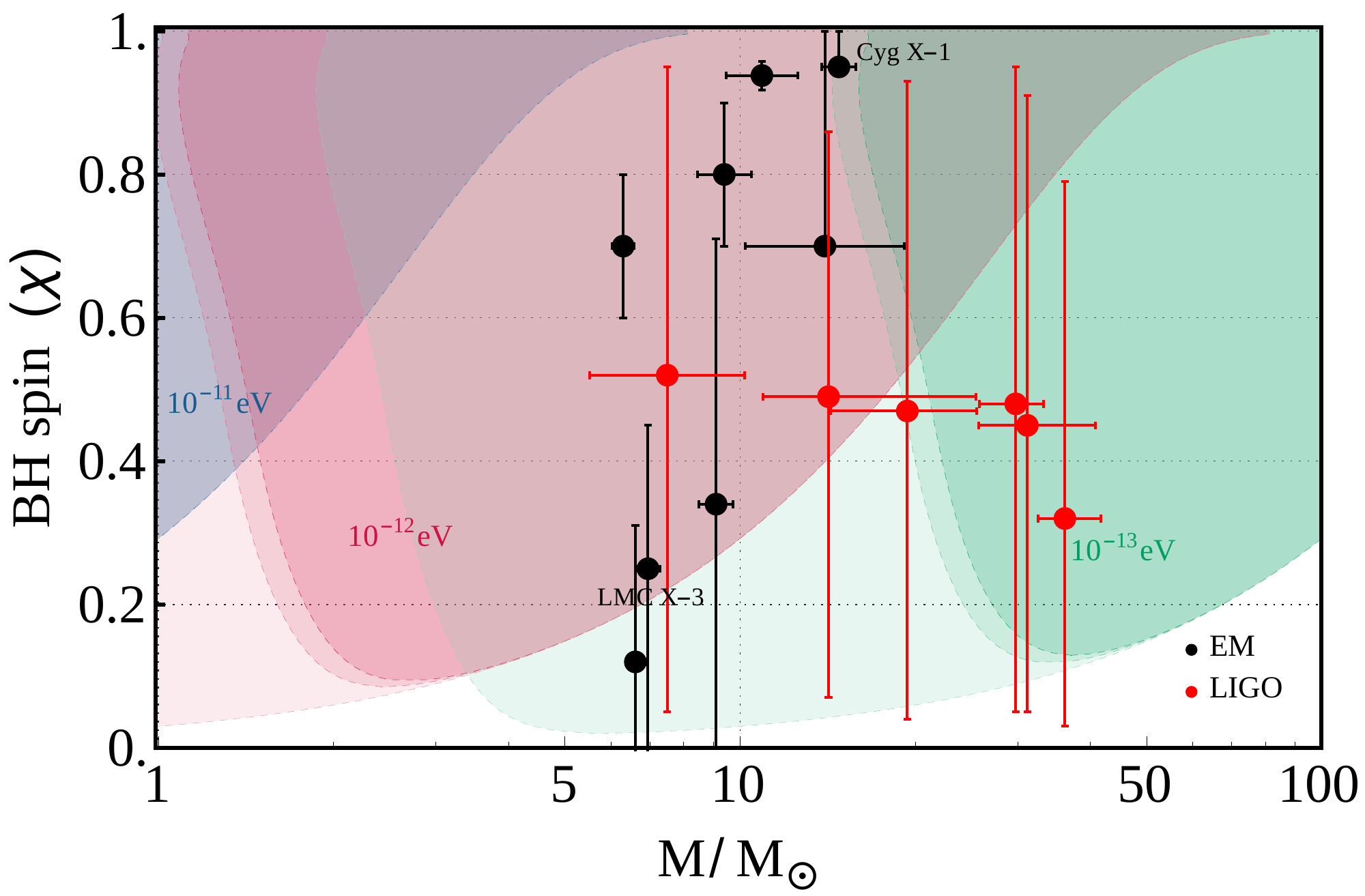}
\includegraphics[width=0.47\textwidth]{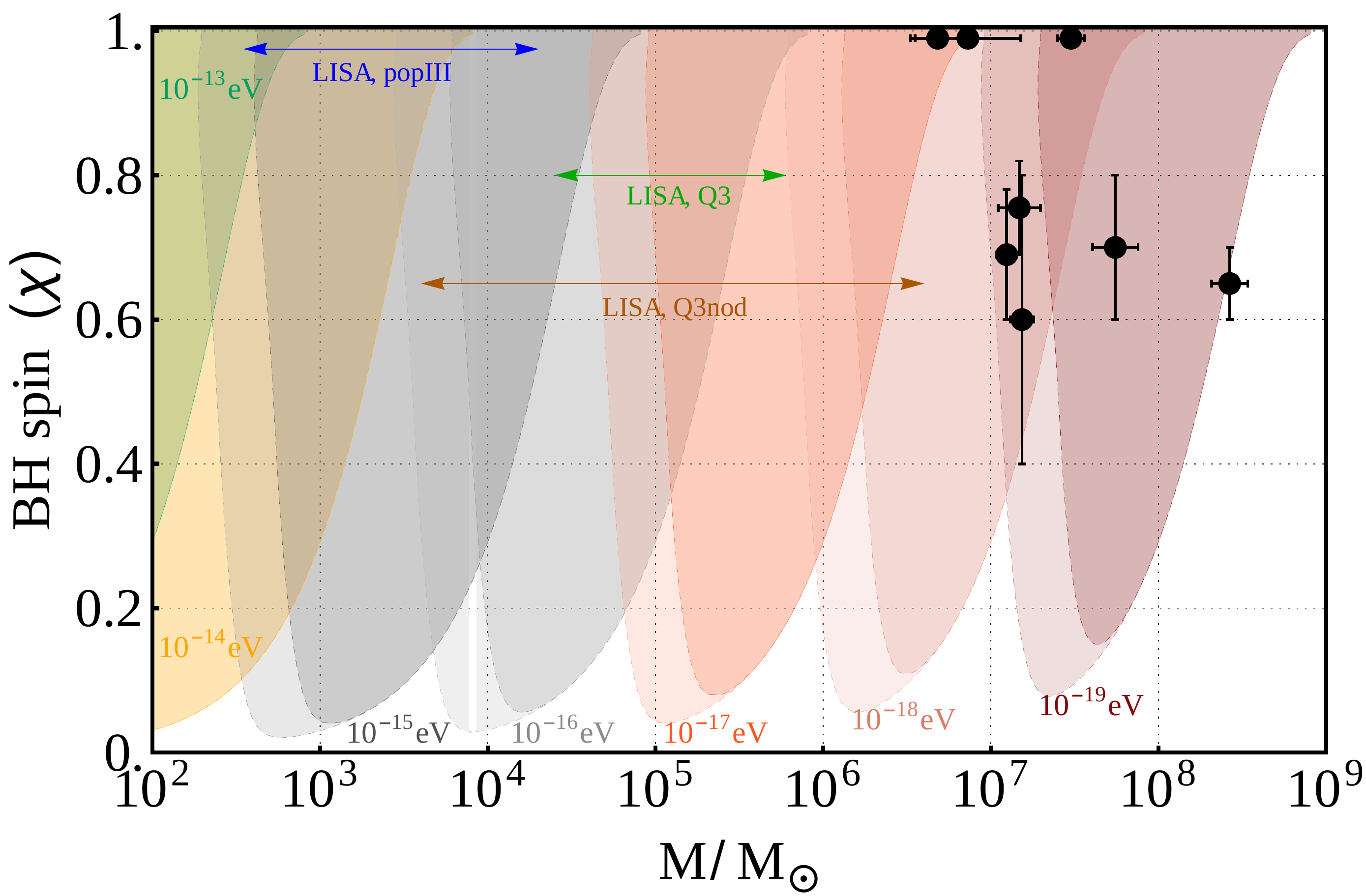}
\caption{Exclusion regions in the BH spin-mass diagram obtained from the superradiant instability of Kerr BHs against massive vector (Proca) fields. 
Left panel: for each Proca mass, BHs laying in the shaded region would have an instability time scale shorter than $14\,{\rm yr}$ (darker filling), $26\,{\rm yr}$ (lighter filling) or $5\,{\rm Myr}$ (even lighter filling).
Right panel: for each Proca mass, BHs laying in the shaded region would have an instability time scale shorter than $5\,{\rm Myr}$ (darker filling) or $500\,{\rm Myr}$ (lighter filling).
Black markers (with error bars) are mass and spin measurements of supermassive BHs obtained through the K$\alpha$ iron line or by continuum fitting~\cite{Brenneman:2011wz,Middleton:2015osa}. Red markers denote LIGO measurements of the spin of the primary and secondary BH for GW150914, GW151226 and GW170104~\cite{TheLIGOScientific:2016pea,Abbott:2017vtc}. The arrows denote the range of projected LISA measurements using three different population models for supermassive BH growth (popIII, Q3 and Q3-nod from~\cite{Klein:2015hvg}, cf. Ref.~\cite{Brito:2017zvb} for details).
\label{fig:Regge_Proca}}
\end{center}
\end{figure}

Figure~\ref{fig:Regge_Proca} summarizes the constraints on dark photons coming from the mass-spin measurements of stellar (left panel) and supermassive (right panel) BHs. For each Proca mass, BHs laying in the shaded region would have an instability time scale shorter than a given threshold.

As previously discussed, the location of the inner disk of stellar mass BHs is observed to remain 
constant over several years, strongly suggesting that the spin of these objects is constant over (at least) the same period. 
For this reason, in the left panel of Fig.~\ref{fig:Regge_Proca} the separatrix for each Proca mass is defined by $\tau=(14,26)\,{\rm yr}$. These two values correspond to the baseline over which the spin of sources Cyg~X-1 and LMC~X-3 is measured to remain stable ($a/M>0.95$ at $3\sigma$ for Cyg~X-1 and $a/M\approx0.5$ for LMC~X-3~\cite{2010ApJ...718L.117S}, respectively).
Therefore, mass and spin measurements of Cyg~X-1 and LMC~X-3 already put a robust constrain on dark photons. The rest of the observational points in the left panel of Fig.~\ref{fig:Regge_Proca} would imply similar constraints \emph{assuming} their spin is observed to remain constant over such baseline.
For completeness, in the left panel of Fig.~\ref{fig:Regge_Proca} we also consider a larger threshold given by a typical accretion time scale, which is customary when constraining the BH Regge plane~\cite{Arvanitaki:2010sy,Pani:2012vp,Pani:2012bp,Brito:2014wla,Baryakhtar:2017ngi}. We considered $\tau=5 {\rm Myr}$, corresponding to an extremely conservative accretion time scale, obtained from the Salpeter time in Eq.~\eqref{salpeter} by considering an efficiency $\eta\approx0.1$ and $f_{\rm Edd}\approx 10$ to account for phases of super-Eddington accretion. Obviously, the excluded region is larger in this case, although the corresponding constraints are less robust because they assume that the BH spin is constant at the observation time.

The excluded regions become smaller as the Proca mass decreases, which corresponds to BHs of larger mass. However, for supermassive objects it is more reasonable to compare the instability time scale with the typical time scale for accretion. This is done in the right panel of Fig.~\ref{fig:Regge_Proca}, where the separatrix for each Proca mass is defined by $\tau=(5,500)\,{\rm Myr}$. The latter threshold is obtained from Eq.~\eqref{salpeter} with $\eta\approx0.1$ and a more realistic choice $f_{\rm Edd}\approx 0.1$. Furthermore, this time scale was chosen because it also roughly corresponds to the age of the Universe at redshift $z\approx 10$, which will be relevant for LISA detections of supermassive BHs at cosmological distance~\cite{Audley:2017drz} (see Refs.~\cite{Brito:2017wnc,Brito:2017zvb} for a detailed analysis of LISA constraints on light scalar fields).


\subsection{Constraints on ALPs}

%
\begin{figure}[th]
\begin{center}
\includegraphics[width=0.47\textwidth]{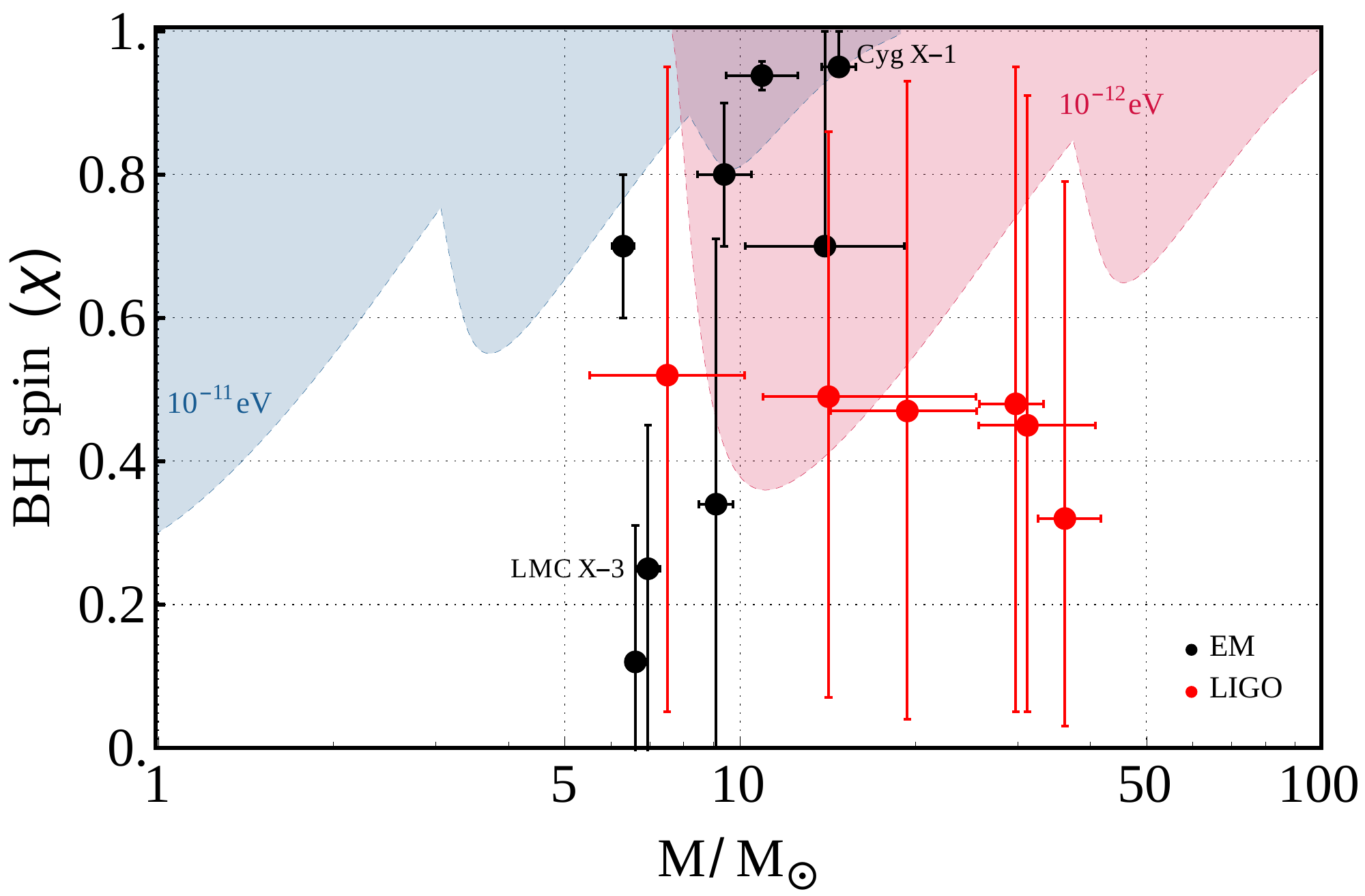}
\includegraphics[width=0.47\textwidth]{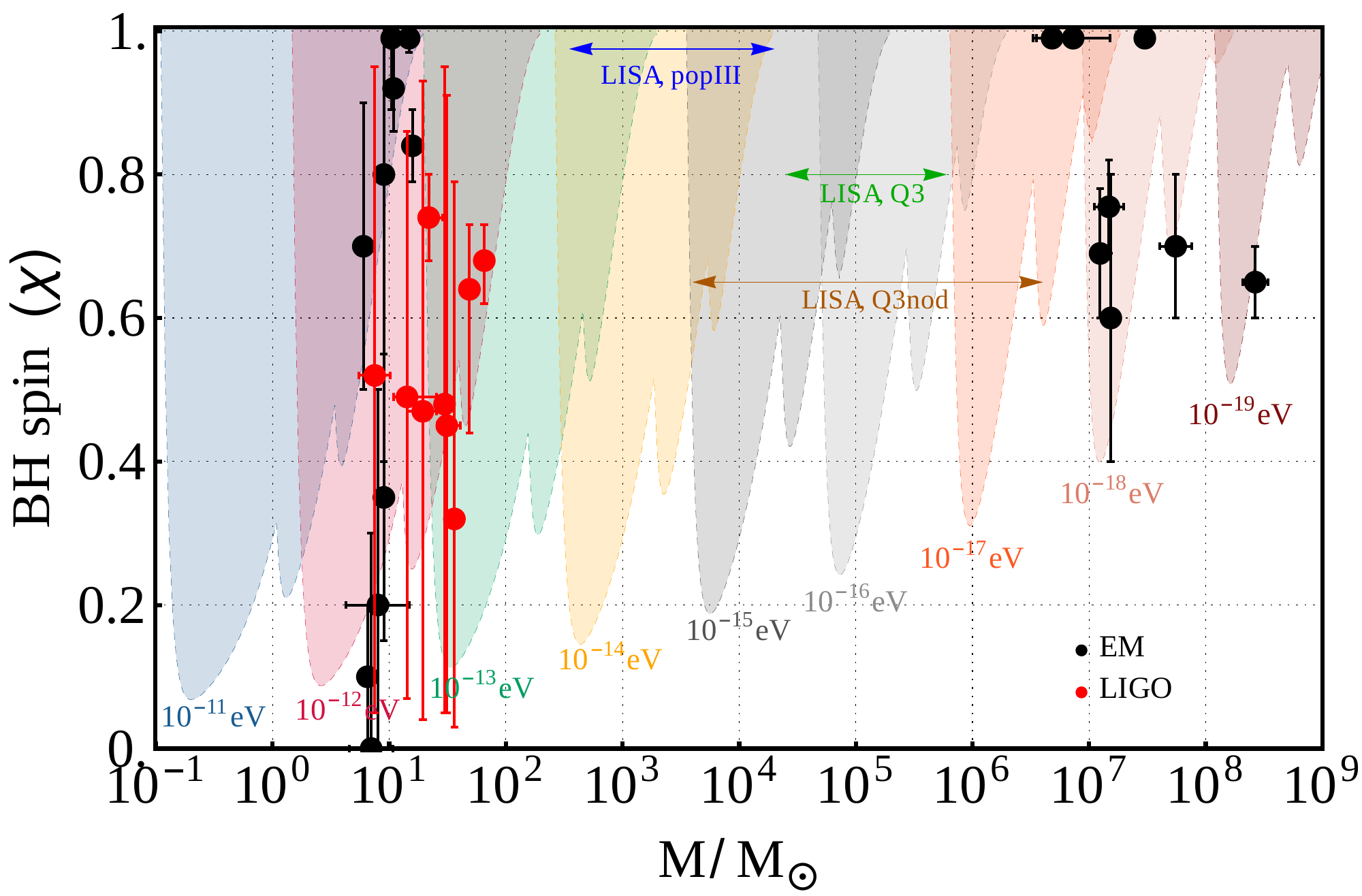}
\caption{Exclusion regions in the BH spin-mass diagram obtained from the superradiant instability of Kerr BHs against massive scalar fields for the most unstable $m=1,2,3$ modes.
For each axion mass, the separatrix in the left (right) plot corresponds to an instability time scale equal to $14\,{\rm yr}$ ($50\,{\rm Myr}$). Markers are the same as those in Fig.~\ref{fig:Regge_Proca}.
\label{fig:Regge_scalar}}
\end{center}
\end{figure}

For completeness, in Fig.~\ref{fig:Regge_scalar} we repeat the analysis previously done for ULVs and we compute the exclusion regions in the BH spin-mass diagram obtained from the superradiant instability of Kerr BHs against massive scalar fields, which was explored in detail in previous work. In this case, we have included also the first most unstable modes with $l=m>1$, namely $l=m=2,3$.

To avoid cluttering of the plots, we have considered only one threshold time scale for each panel: in the left panel we consider the most conservative case, $\tau=14\,{\rm yr}$, corresponding to the baseline of Cyg-X1, whereas in the right panel we considered $\tau=50\,{\rm Myr}$, corresponding to the Salpeter time scale
defined in Eq.~\eqref{salpeter} for a typical efficiency $\eta\approx0.1$ and Eddington mass-accretion rate $f_{\rm Edd}\approx1$.

Comparison between Fig.~\ref{fig:Regge_Proca} and Fig.~\ref{fig:Regge_scalar} leads to two important remarks. First, the constraints on ULVs are more stringent than those on ALPs, because the superradiant instability of the former is more efficient~\cite{Pani:2012vp,Pani:2012bp,Baryakhtar:2017ngi,East:2017ovw}.
Furthermore, higher-$m$ modes (previously neglected for the Proca field) would allow to extend the rightmost part of each Regge gap shown in Fig.~\ref{fig:Regge_Proca}. Thus, the constraints that can be derived from Fig.~\ref{fig:Regge_Proca} are in fact conservative and should be extended by including higher-$m$ Proca modes as done in Ref.~\cite{Baryakhtar:2017ngi} in the Newtonian limit.

\section{Conclusion and discussion}
We have computed, for the first time, the entire spectrum of the most unstable superradiant modes of a Proca field around a Kerr BH. Our results confirm and extend some recent work, showing that the instability of massive vector fields is stronger than in the scalar case. As a consequence, the constraints on the Proca mass are more stringent than those of the scalar mass.
Our results, together with the observed stability of the inner disk of stellar-mass BH candidates, can be used to exclude ULVs and ALPs in the mass range
\begin{eqnarray}
 m_{V}   &\in& (10^{-13}, 3\times 10^{-12})\,{\rm eV}\,,\\
 m_{\rm ALP}  &\in& (6\times 10^{-13}, 10^{-11})\,{\rm eV}\,.
\end{eqnarray}
The lower limit on ALPs is less stringent because the superradiant instability is weaker, whereas the upper limit is more stringent because of the inclusion of higher-$m$ modes.
Note that the above bounds are necessarily less stringent than those obtained by comparing the instability with accretion. Nonetheless, they are much more robust as based on direct observations. To the best of our knowledge, it is the first time that the observed stability of the inner disk of BH candidates is used in the context of BH superradiance.

Statistical evidence of highly-spinning BHs provides an indirect way to constrain ultralight bosons~\cite{Arvanitaki:2014wva,Arvanitaki:2016qwi,Baryakhtar:2017ngi,Brito:2017wnc,Brito:2017zvb} and possibly to measure their mass in case of detection~\cite{Brito:2017zvb}.
In particular, recent work shows that LISA will be able to fill the gap between stellar BHs and supermassive BHs in the Regge plane~\cite{Brito:2017wnc,Brito:2017zvb}. Similarly to the scalar case~\cite{Brito:2017zvb}, our results suggest that LISA will be able to constrain the range
\begin{equation}
 10^{-19}\,{\rm eV}\lesssim m_{{\rm ALP}, V} \lesssim 10^{-13}\,{\rm eV}\,.
\end{equation}
Overall, present and future constraints will allow to probe the entire region $10^{-19}\,{\rm eV}\lesssim m_{{\rm ALP}, V}\lesssim 10^{-11}\,{\rm eV}$, as discussed in Refs.~\cite{Brito:2017wnc,Brito:2017zvb} for ultralight scalars.

An obvious extension of this work is the inclusion of higher-$m$ modes, which would enlarge the exclusion regions shown in Fig.~\ref{fig:Regge_Proca} similarly to those in Fig.~\ref{fig:Regge_scalar}. Other interesting extensions are the inclusion of the gravitational-wave emission for ULVs (extending the analysis of Ref.~\cite{Baryakhtar:2017ngi} beyond the Newtonian regime and along the lines of Refs.~\cite{Brito:2017wnc,Brito:2017zvb} for the scalar case), the study of the evolution of the superradiant instability in the case of real fields (extending the recent numerical analysis of Ref.~\cite{East:2017ovw}), inclusion of plasma effects~\cite{Pani:2013hpa,Conlon:2017hhi} and boson self-interactions~\cite{Yoshino:2012kn}, and extension to massive spin-2 fields, whose instability time scale is even shorter than for spin-1 fields~\cite{Brito:2013wya}. Finally, it would be interesting to investigate the Proca instability of highly-spinning conducting stars, as recently done in the slow-rotation limit~\cite{Cardoso:2017kgn}.

\section{Acknowledgments}
We thank Joerg Jaeckel and Javier Redondo for interesting correspondence and for providing an updated version of Fig.~\ref{fig:boundsbosons}.
O.D.\ is supported by the STFC Ernest Rutherford grants ST/K005391/1 and ST/M004147/1. G.H. is also supported by the latter. O.D. further acknowledges support from the STFC ``Particle Physics Grants Panel (PPGP) 2016", grant ref. ST/P000711/1. 
V.C.\ acknowledges financial support provided under the European Union's H2020 ERC Consolidator Grant ``Matter and strong-field gravity: New frontiers in Einstein's theory'' grant agreement no. MaGRaTh--646597.
Research at Perimeter Institute is supported by the Government of Canada through Industry Canada and by the Province of Ontario through the Ministry of Economic Development $\&$
Innovation.
P.P.\ acknowledges financial support provided under the European Union's H2020 ERC, Starting Grant agreement no.~DarkGRA--757480.
J.E.S.\ was supported in part by STFC grants PHY-1504541 and ST/P000681/1.
This article is based upon work from COST Action CA16104 ``GWverse'', and MP1304 ``NewCompstar'' supported by COST (European Cooperation in Science and Technology).
This work was partially supported by the H2020-MSCA-RISE-2015 Grant No.~StronGrHEP-690904.

\appendix
\section{Numerical procedure to compute the spectrum of Proca fields\label{sec:appendix}}

Taking the external derivative of \eqref{ProcaEOM} requires that for a massive Proca field one must have
\begin{equation}\label{ProcaEOM2}
\mu_{\gamma}^2 \nabla_\nu B^\nu=0.
\end{equation}
Thus, the condition $\nabla_\nu B^\nu=0$ --~that for a massless field is just a gauge choice known as the Lorentz gauge condition~-- is promoted to a constraint that must be obeyed by the field when $\mu_{\gamma}\neq 0$. It gives an algebraic equation for $B_t(r,\theta)$ in terms of the other three components and their first derivatives. 

We will find convenient to work with a compact radial coordinate $y\in[0,1]$ and with a new polar coordinate  $x\in[-1,1]$ related with the standard coordinates $r,\theta$ of \eqref{metric} by
\begin{equation} \label{coordtransf}
r=\frac{r_+}{1-y^2}\,,\qquad \cos\theta=x\sqrt{2-x^2}\,.
\end{equation}
Note that the horizon is located at $y=0$ and asymptotic infinity is at $y=1$. It is further convenient to work with the dimensionless quantities\footnote{Our final results are presented in the main text in terms of the more natural dimensionless quantities $a/M$, $\omega \,M$ and $\mu_{\gamma}\, M$.}  
\begin{equation} \label{dimensionless}
\alpha=\frac{a}{r_+}\,,\quad \widetilde{\omega}=\omega\, r_+\,,\quad \beta=\mu_{\gamma}\, r_+\,.
\end{equation}

In this setting, we have four unknown functions $B_\nu(y,x)$, with $\nu=\{t,y,x,\phi\}$, of two variables. We find that a convenient minimal set of four PDEs to solve for these functions is given by \eqref{ProcaEOM2} and the three components $y,x,\phi$ of \eqref{ProcaEOM}.

These equations have to be solved subject to appropriate physical boundary conditions\footnote{For a detailed and systematic discussion of regularity of perturbations and associated boundary conditions, the reader is invited to see the discussions in \cite{Dias:2010eu,Dias:2010maa} and in the review \cite{Dias:2015nua}.}. We are interested on searching for unstable modes. These have frequencies whose real part is smaller than the potential barrier height set by the Proca field mass, $\omega_R< \mu_{\gamma}$. A Frobenius analysis at the asymptotic infinity $y=1$ then indicates that unstable modes must decay as
\begin{eqnarray}\label{BC:y1}
 && B_{\nu}\big|_{y\to 1}\sim e^{-\frac{\sqrt{\beta^2-\widetilde{\omega}^2}}{1-y^2}}(1-y^2)^{1+\sigma}\:\:\: \hbox{if}\:\: \nu=t,y\,,\quad  B_{\nu}\big|_{y\to 1}\sim e^{-\frac{\sqrt{\beta^2-\widetilde{\omega}^2}}{1-y^2}}(1-y^2)^{\sigma}\:\:\: \hbox{if}\:\: \nu=x,\phi\,,\nonumber\\
 && \hbox{with}\quad  \sigma\equiv i\,(1+\alpha^2)\,\frac{(\beta^2-2\widetilde{\omega}^2)}{2\sqrt{\beta^2-\widetilde{\omega}^2}}.
 \end{eqnarray}
 Here, as a boundary condition, we have already eliminated a solution that grows unbounded at infinity as $e^{\sqrt{\beta^2-\widetilde{\omega}^2}/(1-y^2)}$.  

At the horizon, regularity of the perturbation in ingoing Eddington-Finkelstein coordinates requires that we impose the boundary condition,
\begin{equation}\label{BC:y0}
B_{\nu}\big|_{y\to 0}\sim y^{-2\,i\,\frac{\omega-m\Omega_H}{4\pi T_H}}\quad \hbox{if}\:\: \nu=t,x,\phi\,,\qquad B_{y}\big|_{y\to 0}\sim y^{-2-2\,i\,\frac{\omega-m\Omega_H}{4\pi T_H}}\,,
\end{equation}
which effectively excludes outgoing modes, $\sim y^{2\,i\,(\omega-m\Omega_H)/(4\pi T_H)}$, at the horizon.

Finally, we have to discuss the boundary conditions at north and south poles of the $S^2$. Here, regularity of the perturbations requires that $m$ is quantized to be an integer. We are interested on unstable modes which must have $m\neq 0$. In these conditions, regularity requires that the perturbations behave as 
\begin{equation}\label{BC:x}
B_{\nu}\big|_{x\to \pm 1}\sim (1-x^2)^{|m|},\quad \hbox{if}\:\: \nu=t,x,\phi\,,\qquad  B_{y}\big|_{x\to \pm 1}\sim (1-x^2)^{|m|-1}\,,
\end{equation}
which eliminates irregular modes that would diverge as $(1-x^2)^{-|m|}$.

To impose the boundary conditions \eqref{BC:y1}-\eqref{BC:x} we find convenient to define the new functions $q_i$, $i=1,2,3,4$ defined has
\begin{eqnarray}\label{def:qs}
&& B_{t}(y,x)= e^{-\frac{\sqrt{\beta^2-\widetilde{\omega}^2}}{1-y^2}}(1-y^2)^{1+ i\,(1+\alpha^2)\,\frac{(\beta^2-2\widetilde{\omega}^2)}{2\sqrt{\beta^2-\widetilde{\omega}^2}}}y^{-2\,i\,\frac{\omega-m\Omega_H}{4\pi T_H}}(1-x^2)^{|m|} \,q_1(y,x),\nonumber\\
&& B_{y}(y,x)= e^{-\frac{\sqrt{\beta^2-\widetilde{\omega}^2}}{1-y^2}}(1-y^2)^{1+ i\,(1+\alpha^2)\,\frac{(\beta^2-2\widetilde{\omega}^2)}{2\sqrt{\beta^2-\widetilde{\omega}^2}}}y^{-2-2\,i\,\frac{\omega-m\Omega_H}{4\pi T_H}}(1-x^2)^{|m|} \,q_2(y,x),\nonumber\\
&& B_{x}(y,x)= e^{-\frac{\sqrt{\beta^2-\widetilde{\omega}^2}}{1-y^2}}(1-y^2)^{i\,(1+\alpha^2)\,\frac{(\beta^2-2\widetilde{\omega}^2)}{2\sqrt{\beta^2-\widetilde{\omega}^2}}}y^{-2\,i\,\frac{\omega-m\Omega_H}{4\pi T_H}}(1-x^2)^{|m|-1} \,q_3(y,x),\nonumber\\
&& B_{x}(y,x)= e^{-\frac{\sqrt{\beta^2-\widetilde{\omega}^2}}{1-y^2}}(1-y^2)^{i\,(1+\alpha^2)\,\frac{(\beta^2-2\widetilde{\omega}^2)}{2\sqrt{\beta^2-\widetilde{\omega}^2}}}y^{-2\,i\,\frac{\omega-m\Omega_H}{4\pi T_H}}(1-x^2)^{|m|} \,q_4(y,x),
\end{eqnarray}
This redefinition of functions together with the coordinate transformations \eqref{coordtransf} have the added value of simplifying considerably the procedure of imposing the boundary conditions discussed above. 
Indeed, in terms of the new variables,  the horizon the boundary conditions are simply 
\begin{equation}\label{BCqs:y0}
q_1(0,x)-\frac{1-\alpha^2}{1+\alpha^2}\,q_2(0,x)+\frac{\alpha}{1+\alpha^2}\,q_4(0,x)=0,\qquad \partial_y q_{2,3,4}(0,x)=0\,,
\end{equation}
while at the poles $x=\pm 1$ the boundary conditions \eqref{BC:x} translate into 
\begin{equation}\label{BCqs:x}
\partial_x q_{1,2}(y,\pm 1)=0\,,\qquad \partial_x q_3(y,\pm 1) = \pm q_3(y,\pm 1) \,,\qquad q_4(y,\pm 1) =\mp i\, \frac{|m|}{2 m} \,q_3(y,\pm 1) \,.
\end{equation}
At $y=1$, we have mixed Robin boundary conditions that are not enlightening to display.

To discretize the field equations we use a pseudospectral collocation grid, in the $x$ and $y$ directions, on Gauss-Chebyshev-Lobbato points. The eigenfrequencies and associated eigenvectors are then found using {\it Mathematica}'s built-in routine {\it Eigensystem}. This method has the advantage of finding several modes (i.e., belonging to different sectors of perturbations and with distinct radial overtones) simultaneously. However, to increase the accuracy of our results at a much lower computational cost we then use a powerful numerical procedure which uses the Newton-Raphson root-finding algorithm discussed in detail in section III.C of the review \cite{Dias:2015nua}. 
These numerical methods are very well tested. In particular they are the same that were used to compute the ultraspinning and bar-mode gravitational instabilities of rapidly spinning Myers-Perry BHs \cite{Dias:2009iu,Dias:2010maa,Dias:2010eu,Dias:2010gk,Dias:2011jg,Dias:2014eua}, the near-horizon scalar condensation and superradiant instabilities of BHs \cite{Dias:2010ma,Dias:2011tj}, the gravitational superradiant instability of Kerr-AdS BHs \cite{Dias:2013sdc,Cardoso:2013pza} and the electro-gravitational quasinormal modes of the Kerr-Newman BHs \cite{Dias:2015wqa}.

\section{On the accuracy of the fitting formulas~\eqref{fitProcaR} and~\eqref{fitProcaI}}\label{app:fit}

\begin{figure}[th]
\begin{center}
\includegraphics[width=0.45\textwidth]{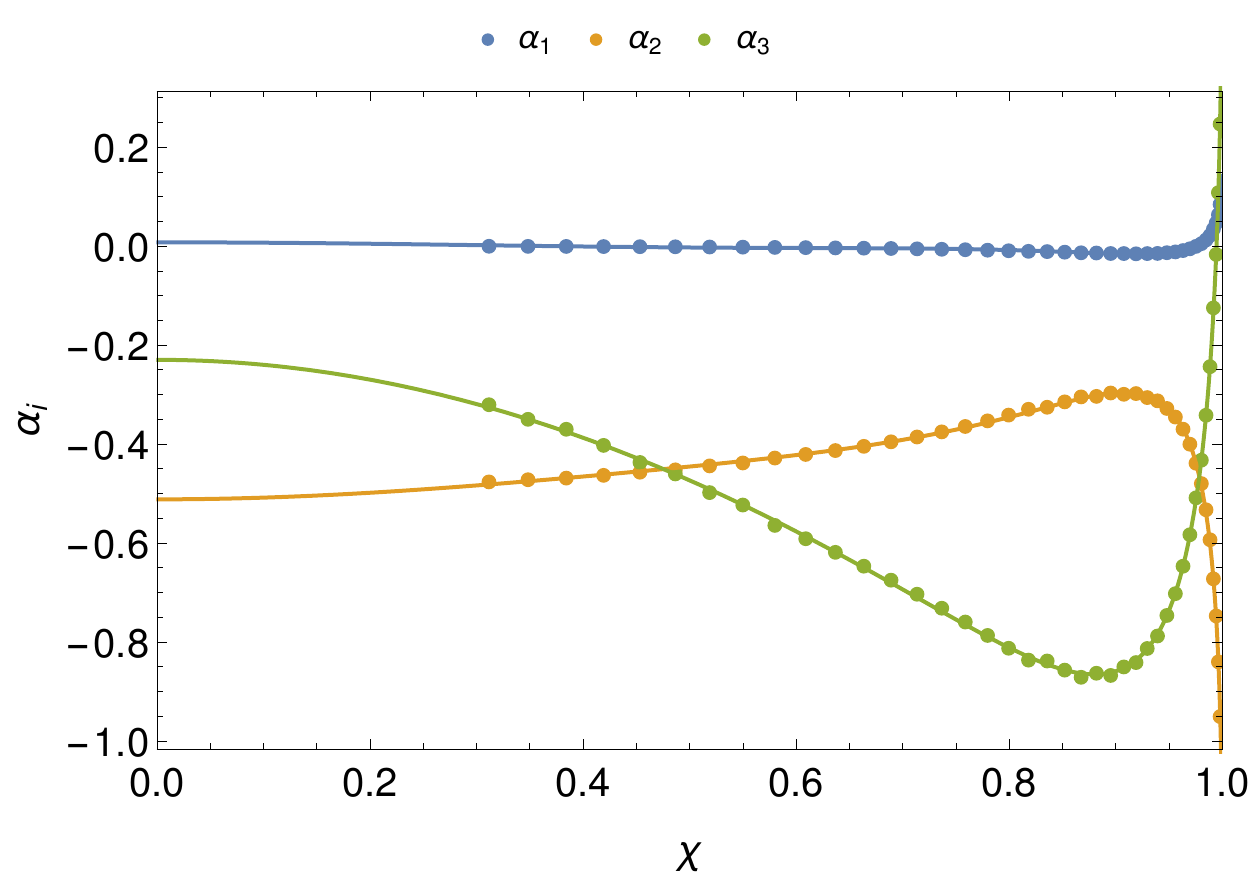}
\includegraphics[width=0.45\textwidth]{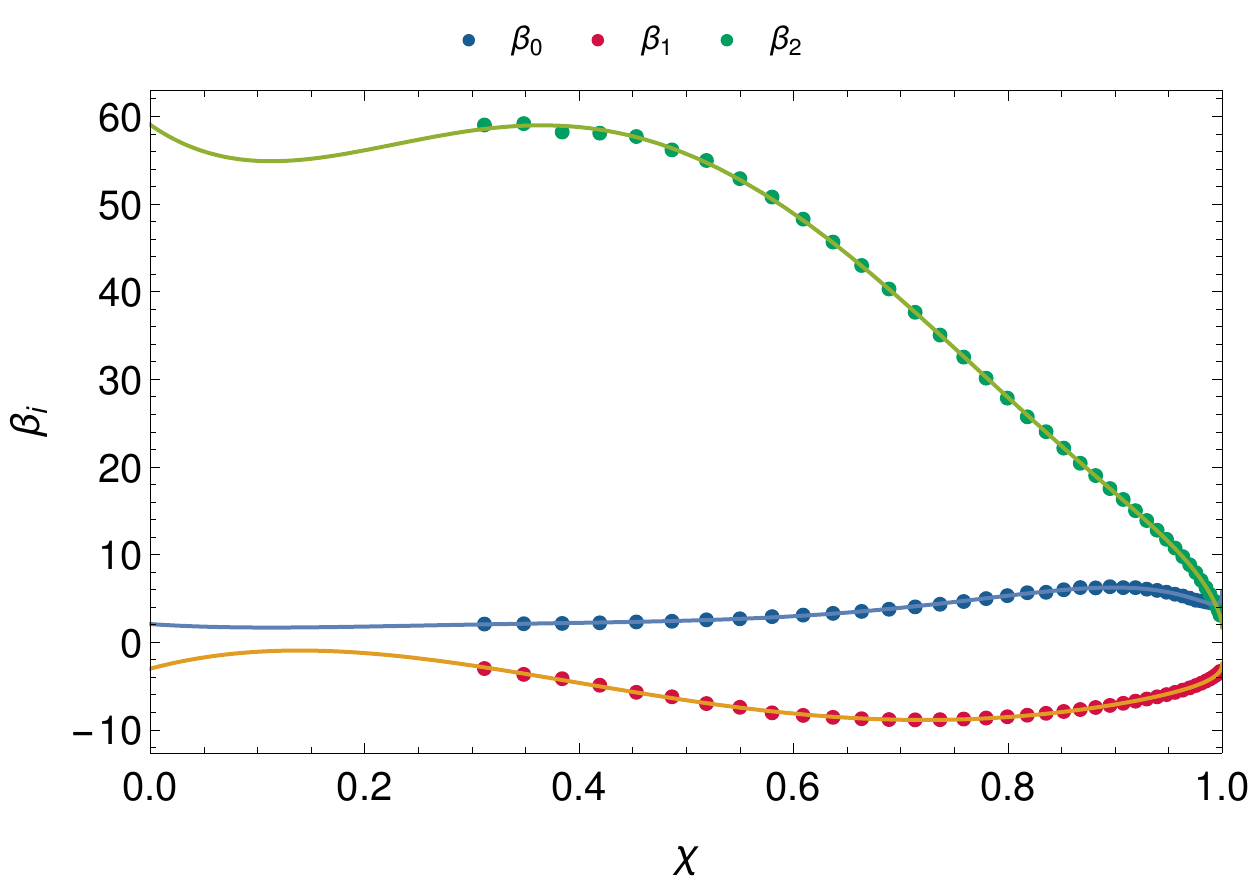}
\caption{The coefficient $\alpha_i$ (left) and $\beta_i$ (right) appearing in Eqs.~\eqref{fitProcaR} and~\eqref{fitProcaI} as functions of the spin and compared to their fitting formulas in Eq.~\eqref{alphabetafit} (cf.\ Table~\ref{tab:fit}). Notice that our data are limited to $\chi\gtrsim0.3$, which is close to the spin threshold for the superradiant instability for $M\mu_{\rm gamma}\approx0.06$, the smallest coupling of our dataset.
\label{fig:alphabetafit}}
\end{center}
\end{figure}

The behavior of the fitting coefficients appearing in Eqs.~\eqref{fitProcaR} and~\eqref{fitProcaI} as a function of the spin is shown in Fig.~\ref{fig:alphabetafit}. 
Interestingly, at variance with what found in Refs.~\cite{East:2017mrj,Baryakhtar:2017ngi}, the subleading terms in Eq.~\eqref{fitProcaR} clearly display some spin dependence at large spin. This includes the coefficient $\alpha_1$, which is almost zero for small spins.
The behavior of $\alpha_i$ in the near extremal limit shows that the first derivative is diverging, which in turn suggests a fitting polynomial in powers of $\sqrt{1-\chi^2}$, as the one adopted in Eq.~\eqref{alphabetafit}. 
In Fig.~\ref{fig:alphabetafit}, we also compare the numerical coefficients with their corresponding fitting formulas given in Eq.~\eqref{alphabetafit}.
The fitting parameters appearing in Eq.~\eqref{alphabetafit} are given in Table~\ref{tab:fit}.

\begin{table}
\centering
\caption{Fitting parameters appearing in Eq.~\eqref{alphabetafit}.}\label{tab:fit}
 \begin{tabular}{cccccc}
  \hline
  \hline
  & $j=0$ & $j=1$ & $j=2$ & $j=3$& $j=4$\\
 \hline
  $A_j^{(1)}$ & $+0.142$ 	& $-1.170$	& $+3.024$	& $-3.234$	& $+1.244$ \\
  $A_j^{(2)}$ & $-1.298$ 	& $+6.601$	& $-15.21$	& $+14.47$	& $-5.070$ \\
  $A_j^{(3)}$ & $+0.726$ 	& $-8.516$	& $+15.43$	& $-11.15$	& $+3.277$   \\
  $B_j^{(1)}$ & -- 		& $-27.76$	& $+114.9$	& $-311.1$	& $+177.2$   \\
  $B_j^{(2)}$ & -- 		& $-14.05$ 	& $+20.78$	& $-36.84$	& $+58.37$  \\
  $B_j^{(3)}$ & -- 		& $+14.78$ 	& $-4.574$	& $-248.5$	& $+108.1$ \\
  $C_j^{(1)}$ & $+48.86$	& $-8.430$	& $+45.66$	& $-132.8$	& $+52.48$     \\
  $C_j^{(2)}$ & $-31.20$	& $+32.52$	& $-73.50$	& $+161.0$	& $-91.27$  \\
  $C_j^{(3)}$ & $+189.1$	& $-85.32$	& $+388.6$	& $-1057$	& $+566.1$       \\
  \hline
  \hline
  \end{tabular}
\end{table}

\begin{figure}[th]
\begin{center}
\includegraphics[width=0.45\textwidth]{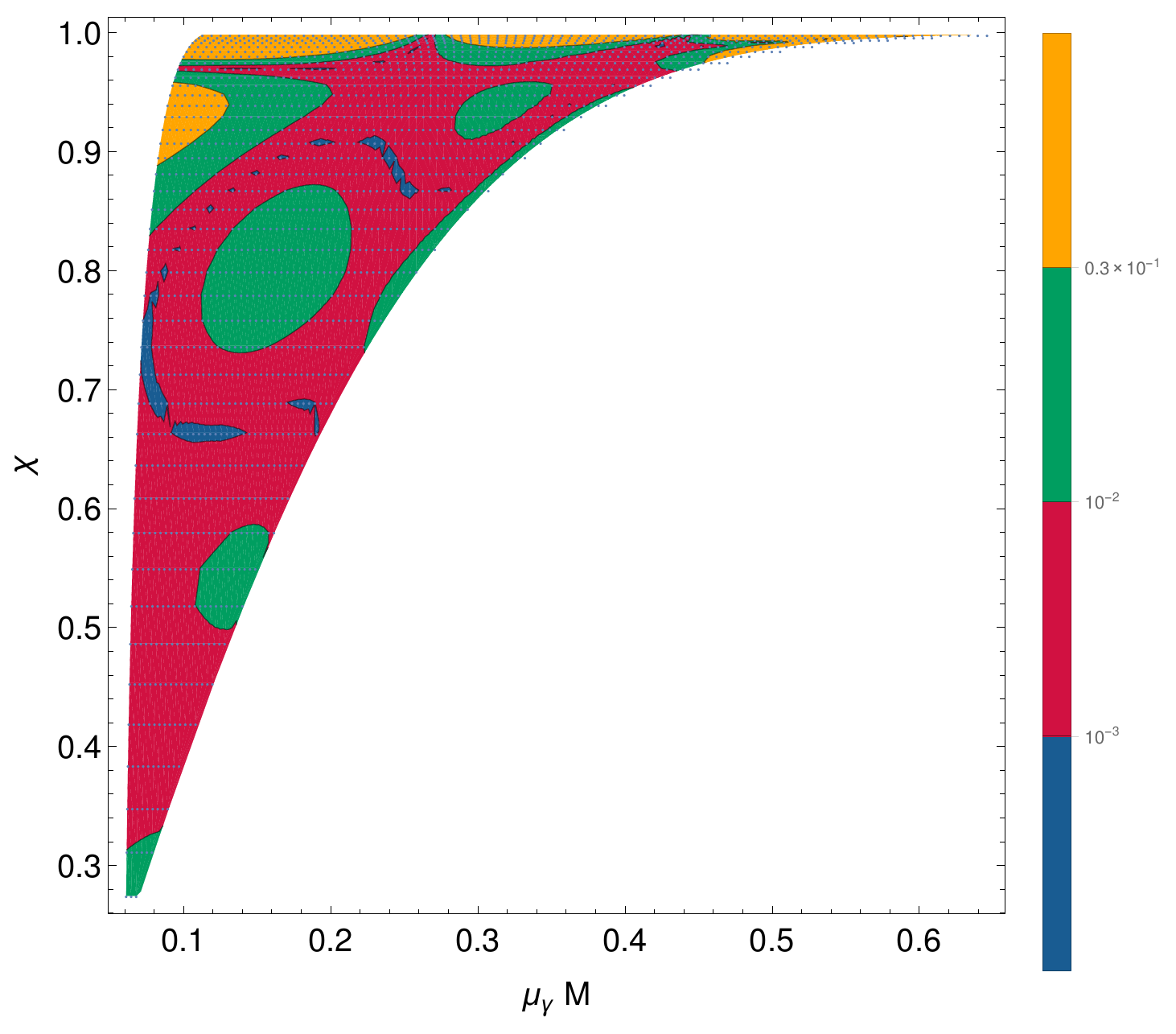}
\includegraphics[width=0.43\textwidth]{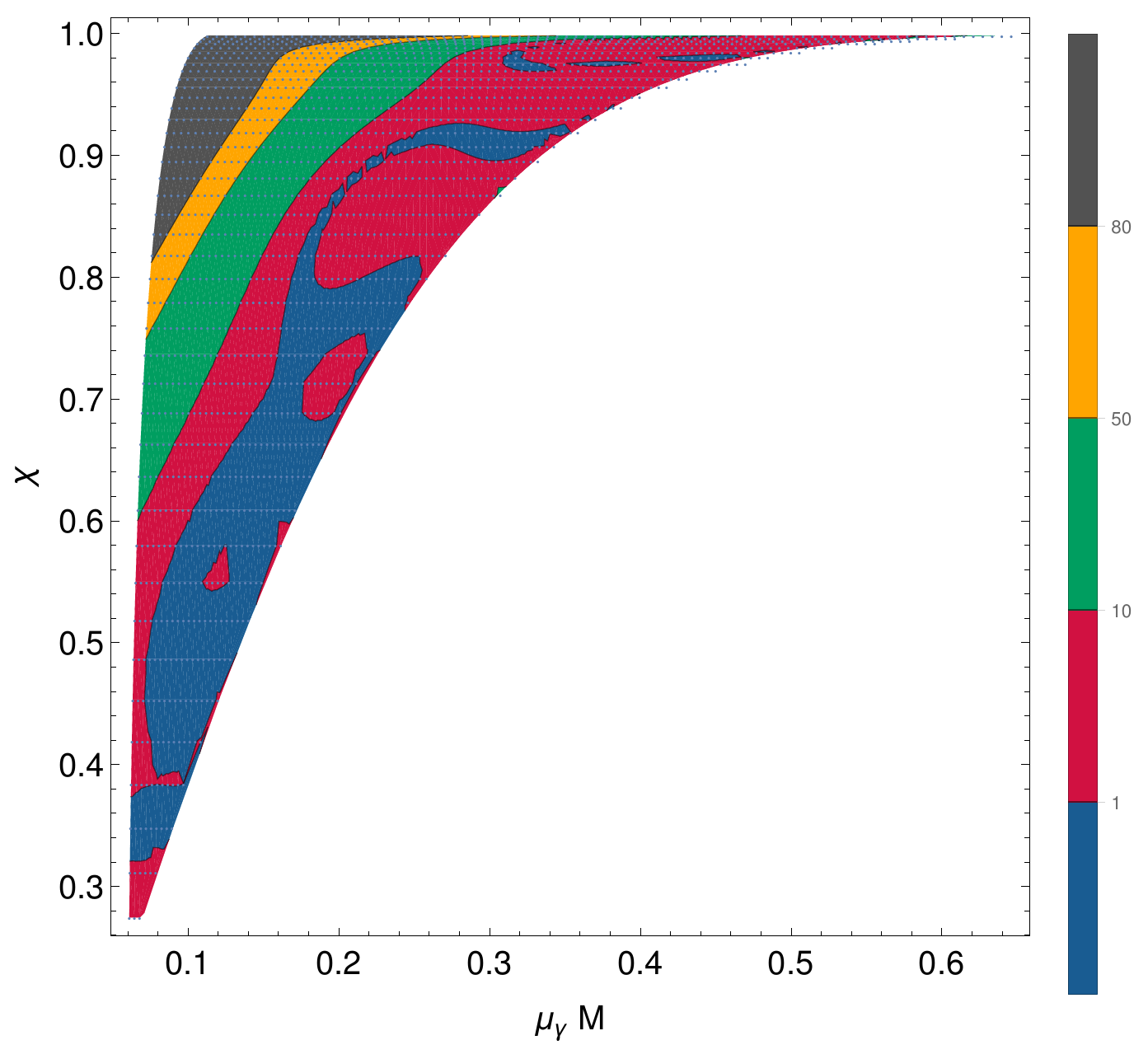}
\caption{Percentage difference between our exact numerical result and the fitting formula~\eqref{fitProcaR} (left) and~\eqref{fitProcaI} (right) in the superradiant region $\omega_R<m\Omega_H$ for the most unstable Proca mode with $m=1$. We also superimpose the values of $(M\mu_\gamma,\chi)$ corresponding to our dataset.
\label{fig:contour}}
\end{center}
\end{figure}

Finally, in Fig.~\ref{fig:contour} we show a contour plot of the (percentage) difference between our exact numerical result and the fitting formulas~\eqref{fitProcaR} and~\eqref{fitProcaI}. The agreement for $\omega_R$ is excellent, whereas the agreement for $\omega_I$ is better than $50\%$ in a large region of the parameter space and it is about a factor of $2$ in the $(a/M\sim1,M\mu_\gamma\sim0)$ corner.
\bibliographystyle{apsrev4}
\bibliography{bib}
\end{document}